\documentclass[%
 reprint,
superscriptaddress,
 amsmath,amssymb,
 aps,
pra,
]{revtex4-2}

\usepackage{graphicx}
\usepackage{dcolumn}
\usepackage{bm}
\usepackage{hyperref}
\hypersetup{colorlinks=true,citecolor=blue,linkcolor=blue,urlcolor=blue}

\newcounter{panel}[figure]

\newcommand{\sublabel}[1]{\refstepcounter{panel}\label{#1}}

\newcommand\identity{1\kern-0.25em\text{l}}

\usepackage{amsmath}
\usepackage{mathtools}

\usepackage{svg}
\usepackage{physics}

\begin{document}

\title{Dissipatively Stabilized 0-n Fock Qubits for Noise-Biased Quantum Computing}

\newcommand{\Caltech}{California Institute of Technology, Pasadena, CA 91125, USA}
\newcommand{\aws}{Amazon Center for Quantum Computing, Pasadena, California 91125, USA}

\author{Su Direkci}
\affiliation{\Caltech}
\author{Simon Lieu}
\affiliation{\aws}
\author{Kyungjoo Noh}
\thanks{Current address: NVIDIA Corporation, USA}
\affiliation{\aws}
\author{Connor T.~Hann}
\affiliation{\aws}

\date{\today}

\begin{abstract}
    Noise-biased qubits have bit-flip errors that are exponentially suppressed relative to phase-flip errors, and offer a promising route toward fault-tolerant quantum computing. However, this bias can be compromised during gate operations with non-biased control qubits. To address this limitation, we propose a ``0-n" Fock qubit architecture that maintains the noise bias by encoding information in the ground state and n-th excited state of a nonlinear multi-level system, such as a transmon. This encoding is achieved via a dissipative stabilization that acts as decay and gain for lower and upper intermediate levels, respectively. We first analytically demonstrate that bit-flip errors are exponentially suppressed with the number of levels. Then, we present a practical implementation using a multi-mode lossy filter to achieve the frequency-selective dissipation. 
    Finally, we numerically demonstrate that bit-flip probabilities approaching $10^{-8}$ are achievable for controlled-X gates on cat qubits using the 0-n qubit as an ancilla with $n \geq 9$ (i.e. ten or more levels), for realistic experimental parameters. Building on this, we simulate syndrome extraction in a repetition code, achieving logical error rates in the megaquop regime with only a distance of $d=9$.
\end{abstract}

\maketitle


\section{Introduction}

Noise-biased physical qubits provide a promising route toward hardware-efficient fault-tolerant quantum computing~\cite{Aliferis2008,Tuckett2018,Tuckett2020, Guillaud2023Review}. When one error channel is strongly suppressed relative to another, this asymmetry can be exploited to achieve improved thresholds and reduced overhead compared to unbiased architectures~\cite{puri_bias-preserving_2020,Webster2015}. Among noise-biased qubits, bosonic encodings have emerged as a platform in which such bias can be engineered at the hardware level~\cite{mirrahimi_dynamically_2014,Michael2016}. In particular, dissipatively stabilized cat qubits—logical states encoded in superpositions of coherent states \cite{Cochrane1999CatCode} and protected by engineered dissipation—have demonstrated exponential suppression of bit-flip errors with increasing mean photon number, $|\alpha|^2$~\cite{lescanne_exponential_2020,putterman_preserving_2025, Berdou2023BitFlipTime, Reglade2024}. Related noise-biased bosonic encodings include Kerr-cat qubits \cite{puri_engineering_2017, grimm_stabilization_2020}, squeezed-cat qubits \cite{Xu2023, Schlegel2022SqueezedCat, Hillmann2023SqueezedCat}, and pair-cat codes \cite{Albert2019PairCat}.

A central requirement in noise-biased architectures is that gates preserve the bias~\cite{puri_bias-preserving_2020}. In particular, syndrome extraction in error correcting codes requires entangling gates between data and ancilla qubits, during which errors on the ancilla can propagate to the data qubit \cite{Aliferis2008, puri_bias-preserving_2020}. If the ancilla is not itself noise-biased, these back-action errors can introduce bit flips on the data qubit and compromise the bias that the encoding was designed to provide. The two potential pathways to address this are i) using cat codes for ancillary qubits, then correcting propagating ancilla errors with an outer error correcting code, and ii) implementing hybrid protocols, where e.g. transmons are used as ancillary qubits~\cite{hann_hybrid_2025}. For the former approach, although significant progress has been made for efficient implementations of bias-preserving gates between cat qubits~\cite{puri_bias-preserving_2020,xu_engineering_2022, Gautier2022CombinedConfinement, Gautier2023ZenoGates}, the associated Hamiltonian engineering and coherence requirements remain nontrivial. In contrast, the latter approach benefits from less stringent experimental requirements. However, transmons can introduce additional decay and dephasing to the data qubits and thereby limit the achievable bias~\cite{Ding2025,Sivak2024}. These considerations motivate alternative noise-biased qubit realizations whose protection mechanism is compatible with simpler couplings and modular device integration.

To this end, we propose a new type of noise-biased qubit, which we refer to as the \emph{0-n Fock qubit}. Here, logical states are encoded in the Fock states $\ket{0}$ and $\ket{n}$ of an anharmonic oscillator, such as a single transmon operated at a suitably large $E_J/E_C$ ratio, or a Kerr nonlinear oscillator. The key feature of this encoding is protection against single-photon loss: a photon-loss event takes $\ket{n}$ to $\ket{n-1}$, outside of the logical subspace, rather than directly inducing a logical bit flip. At least $\lceil n/2 \rceil $ such events must occur to produce a bit-flip $\ket{1_L} \leftrightarrow \ket{0_L}$, which can be suppressed with increasing number of levels in the qubit, analogous to the suppression of bit-flips with increasing $|\alpha|^2$ in cat codes~\cite{lescanne_exponential_2020}. By engineering dissipation that selectively removes population from unwanted intermediate Fock levels while leaving the states $\{\ket{0}, \ket{n} \}$ invariant, the system can be stabilized within the code manifold. This approach builds on established concepts in bosonic error correction and reservoir engineering, where tailored dissipation is used to confine dynamics to protected subspaces~\cite{Poyatos1996,gertler_protecting_2021,Xu2023,Aliferis2008,Leghtas2015TwoPhotonLoss,Touzard2018CoherentOscillations}.

Level-dependent dissipation is necessary to achieve the target qubit structure, which in turn requires nonlinearity to address each transition individually. Frequency-selective dissipation can be implemented by coupling the qubit capacitively to a linear chain of LC resonators that are themselves coupled to a bath mode. The effective spectral density of the bath viewed by the qubit can be shaped by engineering the inter-couplings of the resonators. Similar techniques have been used in the literature for various purposes, and are referred to as bath engineering~\cite{Poyatos1996, Murch2012BathEngineering, Kienzler2015}. Multi-mode on-chip filters and metamaterial resonators have recently been demonstrated as practical tools for bath engineering and leakage suppression in superconducting circuits~\cite{Bronn2015BroadbandFilters, Mirhosseini2018Metamaterials, putterman_preserving_2025,thorbeck_high-fidelity_2024,kim_fast_2024}. In our setting, we observe that a small number ($<10$) of filter modes are sufficient to achieve strong spectral selectivity while maintaining realistic device parameters.

To demonstrate the performance of the 0-n Fock qubit, we consider controlled-X (CX) gates with 0-n Fock qubits as control and stabilized cat qubits as target, enabling syndrome extraction for the repetition code with the Fock qubits as ancillas. These gates can be realized with only the simple, native dispersive coupling between the Fock qubit and the cat qubits (without additional Hamiltonian engineering). We find that bit-flip probabilities below $10^{-8}$ are achievable with realistic experimental considerations for $n=9$ \cite{wang_high-_2025}.
Furthermore, employing a repetition code with the Fock ancilla, we find logical error rates reaching the megaquop regime ($p_L^*=10^{-6}$) at a code distance of $d=9$, in an experimentally realistic parameter regime. These results demonstrate that frequency-selectively stabilized Fock encodings provide a viable and compact route to realizing and scaling-up bias-preserving architectures, and suggest new directions for integrating bosonic qubits with engineered dissipation in scalable superconducting architectures.

This paper is organized as follows. We first introduce the 0-n Fock qubit in Sec.~\ref{sec:0_n_qubit}, and show that it exhibits noise bias, with bit-flip times that scale exponentially with $n$. We then discuss the implementation of the 0-n qubit using the multimode bosonic filter in Sec.~\ref{sec:filter_implementation}. Finally, in Sec.~\ref{sec:0_n_performance}, we analyze its use as an ancilla for cat-qubit syndrome measurements for realistic experimental parameters. 

\section{Properties of the 0-n qubit}
\label{sec:0_n_qubit}

We define the 0-n Fock qubit using the lowest $N=n+1$ Fock levels of an anharmonic oscillator. The logical states correspond to the lowest and the highest of these levels, with $\ket{0_L} = \ket{0}$, $\ket{1_L} = \ket{n}$. A schematic of the qubit can be found in Fig. \ref{fig:intro_fig_a}. For the stabilization of the qubit in the $\{\ket{0}, \ket{n} \}$ manifold, we assume engineered gain (decay) between levels $m$ and $m+1$ for $\lceil \frac{n}{2} \rceil \leq m \leq n-1$ ($0 \leq m \leq \lfloor \frac{n}{2} \rfloor-1$), $m \in \mathbb{N}$. Let us define the engineered dissipation rate as $\kappa_s$. If $n$ is even, the Fock level $\ket{n/2}$ experiences both engineered gain and decay, which may hinder the stabilization. For this purpose, we assume that $n$ is odd for the rest of the Sections. 

In addition to the engineered dissipation $\kappa_s$ in the designated levels mentioned above, the 0-n qubit is also subject to intrinsic single-photon events:  decay ($\kappa_i^\downarrow$) and heating ($\kappa_i^\uparrow$).
The corresponding Lindblad dissipators are $\kappa_i^\downarrow\mathcal{D}[\hat{a}]$ and $\kappa_i^\uparrow\mathcal{D}[\hat{a}^\dagger]$, where $\mathcal{D}[\hat{A}] \coloneqq \hat{A}\hat{\rho} \hat{A}^\dagger - \frac{1}{2} \{\hat{A}^\dagger \hat{A}, \hat{\rho}\}$, and $\hat{a}$ is the annihilation operator associated with the Fock qubit. 
We consider the regime
\begin{align}
\kappa_i^{\uparrow, \downarrow} \ll \kappa_s,
\end{align}
that is, the engineered dissipation rate greatly exceeds the intrinsic decay and heating rates.

To analyze the dynamics of this system, we consider an $N=n+1$ level system with levels $m=0, 1,\dots,n$, and
dissipation between adjacent levels,
we examine how the system evolves under two sets of Lindblad operators describing the decay and gain between the levels respectively, $\{(\kappa^\text{d}_{j})^{1/2} \ket{j}\bra{j+1}\}$ and $\{(\kappa^\text{h}_{j})^{1/2} \ket{j+1}\bra{j}\}$. The transition rates are then given by $\kappa_{j}^{\text{d,h}}$, and we assume that $\kappa_{j}^{\text{d,h}} \geq 0$, $\kappa_{-1}^{\text{d,h}} = \kappa_{n}^{\text{d,h}} = 0$. 
The evolution of the density matrix $\hat{\rho}(t)$ can be described with \cite{BreuerPetruccione2002}
\begin{align}
\label{eq:lindblad_eqn}
    \frac{d\hat{\rho}}{dt} 
    = \mathcal{L}(\hat\rho)&= \sum_{j=0}^{n}  \ket{j} \bra{j}\left( \kappa_{j}^\text{d} \rho_{j+1, j+1} + \kappa_{j-1}^\text{h} \rho_{j-1, j-1} \right) \nonumber \\
    & \quad \; -\frac{1}{2} \{\ket{j} \bra{j}, \left( \kappa_j^\text{h} + \kappa_{j-1}^\text{d} \right)\hat{\rho} \} 
\end{align}
where $\rho_{j,j} \coloneqq \bra{j}\hat{\rho} \ket{j}$. Note that by construction, jump operators of the engineered dissipation annihilate both $\ket{0}$ and $\ket{n}$, so any logical state $c_0|0\rangle + c_1|n\rangle$ is a dark state of the engineered dissipation.

\begin{figure}
    \centering
\includegraphics[width=\linewidth]{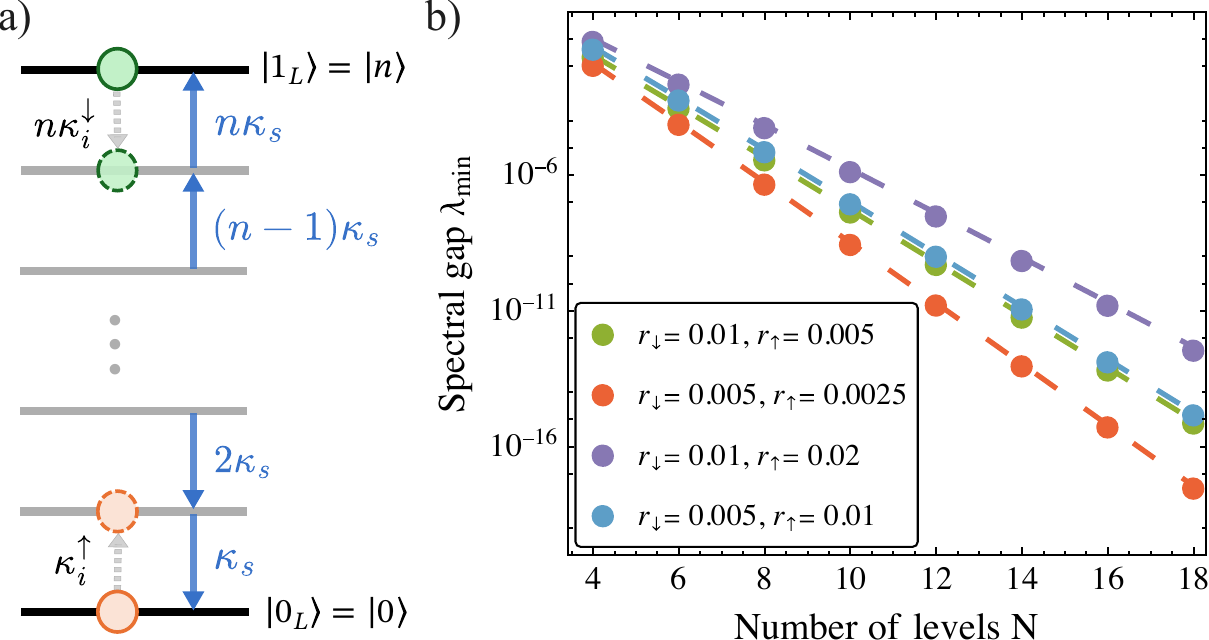}
    \caption{a) Schematic of the 0-n Fock qubit. We encode the logical states $\{\ket{0_L}, \ket{1_L}\}$ into the ground state $\ket{0}$ and the n-th excited state $\ket{n}$ of a Fock qubit. b) Smallest non-zero eigenvalue (in absolute magnitude) $\lambda_\text{min}$ of the Lindbladian describing the evolution of the 0-n Fock qubit. We define $r_{\uparrow, \downarrow} \coloneqq \kappa_i^{\uparrow, \downarrow}/\kappa_s \ll 1$ as the ratio of the rates between the intrinsic processes and the engineered dissipation. The numerical values are plotted with the markers, whereas the analytical fit given by Eq. (\ref{eq:bit_flip_rate_spectral_gap}) is plotted with the dashed lines.}
    \label{fig:intro_fig}
    \sublabel{fig:intro_fig_a}\sublabel{fig:intro_fig_b}
\end{figure}

In Eq. (\ref{eq:lindblad_eqn}), if the initial state is diagonal in the Fock eigenbasis, i.e. $\langle m | \hat{\rho}(0) | n \rangle = 0$ for all $m \neq n$, then the evolved state $\hat{\rho}(t)$ remains diagonal for all $t \ge 0$. Therefore, the evolution for this case reduces to a classical rate equation for the level populations, which can be described in the form of a continuous-time Markov chain (CTMC). In contrast, for an initial coherence $\hat{\rho}(0) = \ket{m}\bra{n}$, $m \neq n$, the state decays with $\hat{\rho}(t) = \exp(-\frac{1}{2}(\kappa_{m}^t+ \kappa_{n}^t)t) \hat{\rho}(0)$, $\kappa_{n}^t = \kappa_{n-1}^\text{d} + \kappa_{n}^\text{h}$ is the total rate of dissipation out of level $n$. Then, the Lindbladian yields the diagonalization $\mathcal{L} = \text{diag}[Q, \mathcal{L}_\text{coh}]$, where $Q$ is a transition matrix describing how the diagonal elements evolve, and $\mathcal{L}_\text{coh}$ encodes the decay of the non-diagonal elements. $\mathcal{L}_\text{coh}$ does not admit steady-state eigenoperators, whereas $Q$ hosts a steady-state solution, characterized by a zero eigenvalue. We will elaborate more on the dynamics of the diagonal and the non-diagonal blocks in Subsections \ref{subsec:noise_bias} and \ref{subsec:phase_coherence}, respectively.

For the 0-n Fock qubit, the transition rates of Eq. (\ref{eq:lindblad_eqn}) are given by, for $ 0 \leq j < n$,
\begin{subequations}
\label{eq:idealized_rates}
\begin{align}
    \frac{\kappa^\text{d}_j}{j+1} &= 
    \begin{cases}
        \kappa_s + \kappa_i^\downarrow & \text{if } j < \lfloor n/2 \rfloor \\
        \kappa_i^\downarrow & \text{otherwise}
    \end{cases} \\[6pt]
    \frac{\kappa^\text{h}_j}{j+1} &= 
    \begin{cases}
        \kappa_s + \kappa_i^\uparrow & \text{if } j \geq \lceil n/2 \rceil \\
        \kappa_i^\uparrow & \text{otherwise}
    \end{cases}
\end{align}
\end{subequations}
Note that the transition rates are multiplied by the bosonic enhancement factor for levels higher than the ground state.

In the following two subsections, we show that the 0-n Fock qubit exhibits a noise bias: bit-flip rates, governed by block $Q$ of the Lindbladian, are suppressed exponentially with $n$ (Subsec. \ref{subsec:noise_bias}); while phase-flip rates, governed by block $\mathcal{L}_\text{coh}$, are enhanced linearly with $n$ (Subsec. \ref{subsec:phase_coherence}). 

\subsection{Evolution of diagonal elements and noise bias}
\label{subsec:noise_bias}

Let us elaborate on the evolution of the diagonal elements of the density matrix. In the CTMC formalism, $\pi(t)$ denotes the population row vector over the $N=n+1$ levels, where $\pi_{j}(t) = \rho_{j,j}(t)$ encodes the population at the $j^\text{th}$ level. Its evolution is governed by
\begin{align}
\pi(t) = \pi(0) e^{Q t}.
\end{align}
Since only the neighboring Fock levels are connected by single-photon processes, $Q$ has a birth-death (tridiagonal) structure. The steady state of the full chain is found from $\pi_{\mathrm{ss}}Q = \mathbf{0}$, where $\mathbf{0}$ is the null row vector. It can be written as
\begin{align}
    \frac{(\pi_{\mathrm{ss}})_{j}}{(\pi_{\mathrm{ss}})_{j+1}} &= \frac{\kappa_i^\downarrow}{\kappa_s + \kappa_i^\uparrow} \approx r_\downarrow, \quad \frac{n+1}{2} \leq j < n \nonumber \\
    \frac{(\pi_{\mathrm{ss}})_{j+1}}{(\pi_{\mathrm{ss}})_{j}} &= \frac{\kappa_i^\uparrow}{\kappa_s + \kappa_i^\downarrow} \approx r_\uparrow, \quad 0 \leq j < \frac{n-1}{2}, \nonumber \\
    \frac{(\pi_{\mathrm{ss}})_{(n+1)/2}}{(\pi_{\mathrm{ss}})_{(n-1)/2}} &= \frac{\kappa_i^\uparrow}{\kappa_i^\downarrow}.
\end{align}
where we define $r_{\uparrow, \downarrow} \coloneqq \kappa_i^{\uparrow, \downarrow}/\kappa_s \ll 1$. This state satisfies the detailed balance equations (i.e. the rate of the ingoing excitations is equal to the rate of the outgoing excitations for each level) \cite{vanKampen2007}, and the relative magnitudes of $\kappa_i^\downarrow$ and $\kappa_i^\uparrow$ determine whether most of the population is concentrated in the upper or lower levels of the qubit. Therefore, due to the intrinsic losses, the qubit cannot be stabilized perfectly to the code manifold. If it is initialized in the highest excited level, $\ket{n}\bra{n}$, it will eventually decay to all of the lower levels, giving rise to a finite bit-flip probability. However, we can inquire how the timescale for the stabilization to the steady state scales as a function of $n$, in order to examine whether the increased number of levels provides a protection from bit-flips.


To analyze this, note that for odd $n$, the total number of levels $N$ is even,
and the Hilbert space separates into two halves connected only by weak intrinsic decay and heating processes. On short timescales, $t \ll 1/\kappa_i^{\uparrow, \downarrow}$, the system rapidly relaxes within each half-space to a metastable distribution governed by the ratio of the intrinsic decay or heating rate to the engineered dissipation rate. Initializing the state in $\hat{\rho} = \ket{n}\bra{n}$, the corresponding metastable state (for upper levels) will be given by
\begin{align}
\label{eq:upper_metastable_state}
\pi_{\mathrm{ms}}^\uparrow \approx \frac{1}{\sum_{j=0}^{\frac{n-1}{2}} \left(r_\downarrow\right)^j} \begin{bmatrix}
0,\dots,0,\left(r_\downarrow\right)^{\frac{n-1}{2}},\dots,r_\downarrow,1
\end{bmatrix},
\end{align}
obtained from detailed balance between neighboring levels. We can also define the metastable state associated with the lower levels as $\pi_\text{ms}^\downarrow$, where the ratio of populations between two levels is similarly given by $r_\uparrow$, in the lower half of the qubit. 

We define a bit-flip as the transfer of population from the upper half ($\frac{n+1}{2} \leq m \leq n$) to the lower half ($0 \leq m \leq \frac{n-1}{2}$) of the qubit. The leakage rate out of this metastable state is, due to intrinsic decay, given by $\frac{n+1}{2} \kappa_i^\downarrow$, where $\frac{n+1}{2}$ is the bosonic enhancement factor. We can similarly write the metastable state corresponding to the lower half of the qubit, where the leakage rate to the upper half would be given by $\frac{n+1}{2} \kappa_i^\uparrow$. 
When population leaks to the opposite half-space, the system quickly relaxes to the corresponding metastable state in time $t \sim O(1/\kappa_s)$. Then, the population vector at some time $t$ can be approximated as $\pi(t) \approx (1-x) \pi_\text{ms}^\uparrow + x \pi_\text{ms}^\downarrow$. The bit-flip probability at this time is therefore $p_\text{bf} = \sum_{j=0}^{(n-1)/2} \pi_j(t) \approx x$. The evolution of $\pi(t)$ is given by $\frac{d}{dt}\pi(t) = \pi(t) Q$. Using this expression and the explicit structure of $Q$, the rate-of-change of the bit-flip probability is found as
\begin{align}
    \frac{d p_\text{bf}}{dt} &\approx (1-p_\text{bf})\sum_{j=0}^{\frac{n-1}{2}} (\pi_\text{ms}^\uparrow Q)_j + p_{\text{bf}} \sum_{j=0}^{\frac{n-1}{2}} (\pi_\text{ms}^\downarrow Q)_j \nonumber \\ &=\frac{(n+1)\kappa_s}{2} \left[ (1-p_\text{bf}) (r_\downarrow)^{\frac{n+1}{2}} - p_\text{bf}\,(r_\uparrow)^{\frac{n+1}{2}}  \right].
\end{align}
Together with the initial condition $p_\text{bf}(0)=0$, the bit-flip probability is therefore
\begin{align}
p_{\mathrm{bf}}(t) &\approx
\frac{1 - \text{exp}\left(-\kappa_s t \frac{n+1}{2}\left[\left(r_\downarrow\right)^{\frac{n+1}{2}}+\left(r_\uparrow\right)^{\frac{n+1}{2}}\right]\right)}{1+\left(r_\uparrow/r_\downarrow\right)^{\frac{n+1}{2}}}.
\end{align}
In the limit of a small exponent, $p_{\mathrm{bf}}(t) \approx \kappa_s t \frac{n+1}{2} \left(r_\downarrow\right)^{\frac{n+1}{2}}$. Then, the effective bit-flip rate
\begin{align}
\label{eq:bit_flip_rate_spectral_gap}
\lambda_\text{bf} = \kappa_s \frac{n+1}{2}\left[\left(r_\downarrow\right)^{\frac{n+1}{2}}+\left(r_\uparrow\right)^{\frac{n+1}{2}}\right]
\end{align}
decreases exponentially with $\frac{n+1}{2}$, up to a linear factor in $n$. This establishes a hardware-level noise bias: logical bit flips require $O(\lceil n/2 \rceil)$ sequential intrinsic decay events and are exponentially suppressed with increasing $n$.
The 0-n Fock qubit therefore exhibits a noise bias arising from the separation between fast engineered stabilization and weak intrinsic excitation or decay.

The effective bit-flip rate can also be found from the smallest (in magnitude) non-zero eigenvalue of the generator matrix $Q$, i.e. the spectral gap. Physically, the spectral gap corresponds to the slowest relaxation mode of the CTMC. For the case outlined here, this is the transfer of population between the upper and lower halves of the levels, which corresponds to a bit-flip. All faster modes, with rates $\sim\kappa_s$, describe relaxation within each half-space to the respective metastable distributions. 

We plot the spectral gap for different values of $r_{\downarrow,\uparrow}$, as a function of the number of levels $N$ of the 0-n qubit, in Fig. \ref{fig:intro_fig_b}. The numerical samples are plotted with the data points, whereas the analytical rate $\lambda_\text{bf}$ (Eq. (\ref{eq:bit_flip_rate_spectral_gap})) is plotted with the dashed lines in the corresponding colors. From the Figure, we observe that the analytical prediction is a very good fit to the numerics.

\subsection{Evolution of non-diagonal elements and phase coherence}
\label{subsec:phase_coherence}

Following a similar analysis to Sec. \ref{subsec:noise_bias}, we derive the phase-flip rate of the 0-n Fock qubit,
given engineered dissipation $\kappa_s$, intrinsic heating $\kappa_i^\uparrow$, and intrinsic decay $\kappa_i^\downarrow$. For this purpose, we define $\ket{\pm_{0n}}=\frac{1}{\sqrt{2}}(\ket{0}\pm\ket{n})$, and initialize the qubit in $\ket{+_{0n}}$. Given the system state $\hat{\rho}(t)$ at some time, we assume an encoding that maps population in the upper and lower half-spaces to $\ket{n}$ and $\ket{0}$, respectively. Given the encoding, the phase-flip probability is given by $p_\text{pf}(t) \coloneqq 1 - \bra{+_{0n}}\hat{\rho}\ket{+_{0n}}$.

From Eq. (\ref{eq:lindblad_eqn}), we observe that off-diagonal terms $\ket{0}\bra{n}$ and $\ket{n}\bra{0}$ will decay with rate $\frac{1}{2}(n \kappa_i^\downarrow + \kappa_i^\uparrow)$. Thus, the phase flip probability is found as
\begin{align}
\label{eq:phase_flip_prob}
    p_\text{pf}(t) &= \frac{1}{2} \left( 1 - \text{exp}\left(-\frac{t}{2}[n \kappa_i^\downarrow + \kappa_i^\uparrow] \right)\right).
\end{align}
$p_\text{pf}(t) \approx \frac{n\kappa_i^\downarrow t}{4}$ for $\frac{t}{2}[n \kappa_i^\downarrow + \kappa_i^\uparrow] \ll 1$, $\kappa_i^\uparrow \sim \kappa_i^\downarrow$, and $n \gg 1$.
This probability grows linearly with $n$. Furthermore, since phase coherence is only affected by heating to levels above $\ket{0}$, or decay to levels below $\ket{n}$, it is not affected by engineered dissipation. This is also demonstrated by the independence of this expression from the engineered dissipation rate, $\kappa_s$. 

In contrast with the exponential suppression of bit-flips, phase-flips are enhanced linearly with the number of levels in the 0-n Fock qubit. This scaling is directly analogous to stabilized cat qubits, which exhibit exponentially suppressed bit flips and linearly enhanced phase flips as the size of the bosonic encoding increases \cite{mirrahimi_dynamically_2014, puri_engineering_2017, lescanne_exponential_2020, grimm_stabilization_2020}. However, different from cat qubits, where the suppression rate of bit-flips is independent of the engineered dissipation rate $\kappa_2$ or the intrinsic dephasing rate $\kappa_\phi$, the suppression rate in the Fock qubit improves as the ratio between the engineered dissipation rate and the intrinsic error rate increases.

\section{Proposed Experimental Implementation}
\label{sec:filter_implementation}


In the previous Section, we have described how the 0-n Fock qubit is noise-biased, assuming that we can engineer the required dissipation structure. In this section, we describe how this dissipation can be realized. The 0-n Fock qubit requires that, in addition to the engineered decay on transitions in the lower half of the level ladder, and engineered gain on transitions in the upper half, each type of engineered dissipation must act only on its designated transitions: unwanted gain on lower levels or decay on upper levels would compromise the stabilization.

To enable this dissipation structure, an anharmonic oscillator such as a Kerr oscillator or a transmon is necessary. In these nonlinear oscillators, each transition $\ket{i} \leftrightarrow \ket{i+1}$ occurs at a distinct frequency, with neighboring transitions separated by the anharmonicity $\eta$. This converts the problem of level-selective dissipation into one of frequency-selective dissipation: if we can engineer a bath whose spectral density has a sharp passband, we can address each transition independently based on its frequency.

To realize such a bath, we couple the qubit to a linear chain of LC resonators terminated by a lossy mode (see Fig. \ref{fig:filter_a}). This chain acts as a bandpass filter, where excitations at frequencies in the passband propagate through the chain and are dissipated, while those outside the passband are reflected. 
As described in detail below, engineered decay and gain can be obtained from two different couplings to the first filter mode: a beam splitter interaction, and a two-mode squeezing interaction, respectively. By choosing the drive frequencies of these interactions, the transitions in the lower half of the qubit can be placed within the decay passband while the upper-half transitions lie outside it (and vice versa for gain). In this way, the single multimode filter simultaneously provides both the frequency-selective decay and the frequency-selective gain required to stabilize the $\{\ket{0}, \ket{n} \}$ manifold.

We first write the dynamics of the multimode filter in Subsec. \ref{subsec:multimode_filter}. The effective dissipation on the Fock qubit is then described in Subsec. \ref{subsec:effective_dissipation}. An implementation of the Fock qubit with the Kerr Hamiltonian is discussed in Subsec. \ref{subsec:kerr_hamiltonian}, and the scalings of the system parameters are derived in Subsec. \ref{subsec:parameter_scaling}.

\subsection{Multimode filter implementation}
\label{subsec:multimode_filter}

Let us first consider engineering decay or gain on a single harmonic oscillator mode $\hat{d}$, with frequency $\omega_d$, before extending to the case of the multi-level anharmonic oscillator. This mode is connected to a linear chain of identical harmonic oscillators $\hat{f}_1, \hat{f}_2, \dots, \hat{f}_M$, with frequency $\omega_f$. The qubit is connected to the first filter mode with (i) a beam splitter interaction, and (ii) a two-mode squeezing interaction. The Hamiltonian in the interaction picture with respect to the bare Hamiltonians of the modes, within the rotating-wave approximation (RWA) can be written as
\begin{align}
\label{eq:qubit_filter_hamiltonian}
    \hat{H} &= g_\text{bs} \, \hat{d} \hat{f}_1^\dagger \, e^{i \Delta_\text{bs} t}  + g_\text{tm}\,\hat{d}^\dagger \hat{f}_1^\dagger \, e^{i \Delta_\text{tm} t} 
    \nonumber \\ & \quad + \sum_{i=1}^{M-1} J_{i,i+1} \, \hat{f}_i \hat{f}_{i+1}^\dagger  + \text{h.c.}
\end{align}
where the first term denotes the beam splitter interaction, second term denotes the two-mode squeezing interaction, and the third term denotes the coupling between adjacent filter modes. $\Delta_\text{bs}$ and $\Delta_\text{tm}$ are the beam splitter and two-mode squeezing parametric drive detunings, respectively. Furthermore, we assume dissipation on the final filter mode, with a dissipator $\kappa_f \mathcal{D}[\hat{f}_M]$. The schematic of this system is plotted in Fig. \ref{fig:filter_a}. 

Such filters have been used extensively in the literature \cite{thorbeck_high-fidelity_2024, chamberland_building_2022, putterman_hardware-efficient_2025, kim_fast_2024} to shape the effective bath seen by the computational qubits, in order to prevent leakage and increase coherence. We describe the resulting dynamics in the following subsection.

\begin{figure}
    \centering
\includegraphics[width=\linewidth]{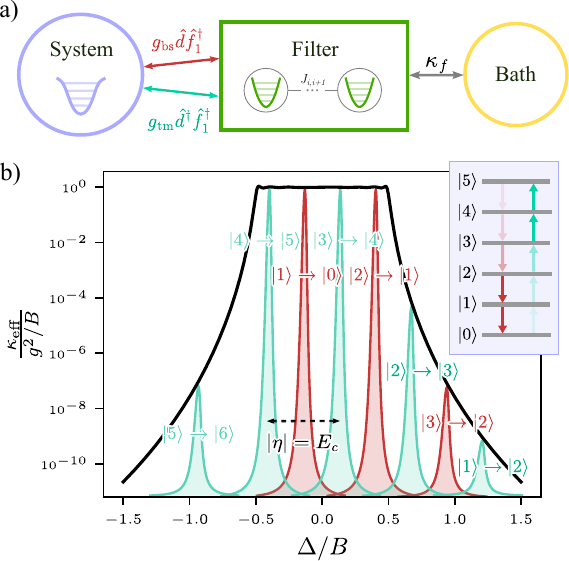}
    \caption{a) Schematic of the implementation for the 0-n Fock qubit. The qubit (system, drawn in purple) is connected to a multimode filter with coupling rates $g_\text{bs}$ and $g_\text{tm}$ for the beam splitter and the two-mode squeezing interaction, respectively. The excitations travel along the filter and are dissipated to a bath, with a rate $\kappa_f$. 
    b) Effective rates $\kappa_\text{eff}$ of engineered processes between adjacent levels of a 0-n Fock qubit, with $n=5$. Decay and gain are engineered via the beam splitter and two-mode squeezing coupling with the filter. The rates are plotted as a function of the detuning $\Delta$, normalized with the filter bandwidth $B$. We observe that $\frac{\kappa_\text{eff}}{g^2/B} \approx 2\pi$, where $g$ is the coupling rate for the corresponding interaction. For an implementation of the Fock qubit with the Kerr Hamiltonian, the difference in detuning between adjacent transitions corresponds to the anharmonicity $\eta$.}
    \label{fig:filter}
    \sublabel{fig:filter_a}
    \sublabel{fig:filter_b}
\end{figure}

\subsection{Effective dissipation}
\label{subsec:effective_dissipation}

In the weak-coupling limit $g_\text{bs}, g_\text{tm} \ll J_i, \kappa_f \; \forall \; i$, the eigenmodes of the coupled system can be obtained perturbatively. For the eigenmode associated with the \textit{bare} system mode, i.e. $\hat{d}$, 
it can be shown that
a two-mode squeezing interaction induces gain while a beam splitter interaction induces decay in the form of $\hat{d}(t) = \hat{d}(0)e^{ -\kappa_\text{eff} t}$, where we refer to $\kappa_{\text{eff}}$ as the effective gain or decay rate. 
Then, by coupling the system mode to a chain of oscillators, we can engineer dissipation that can either inject excitations into or extract excitations from the mode. Moreover, the dissipation profile can be shaped by optimizing over the parameters $\{J_i\}$ and $\kappa_f$.  

Since the dissipation is observed in the eigenmodes of the joint system, i.e. the \textit{dressed} modes that include the system mode and the filter modes, there are some subtleties with how the joint system evolves. We can use adiabatic ramping of the interaction rate $g_{\text{bs, tm}}$ to achieve a smooth transition from the bare eigenmodes to the dressed eigenmodes. If the ramping is slow enough, it can be shown that the effective dynamics on a given mode $\hat{v}$ look like:
\begin{align}
    \hat{v}(t) = \sum_i c_i(0) \exp \left(\int_0^t ds \lambda_i(s) \right) \hat{u}_i(t)
\end{align}
where $\{c_i\}$ are the coefficients determining the decomposition over the eigenmodes $\{\hat{u}_i \}$ at time $t=0$, $\{\lambda_i\}$ are the corresponding eigenvalues, and $\hat{v}(t)$ is the Heisenberg operator of $\hat{v}$. Ramping the interaction rate from zero at $t = 0$ up to $g_{\text{bs, tm}}$ and then down to zero at $t = T$, where $T$ is the total simulation time, results in $c_i(0) = 1$ for $i = i^*$, $c_i(0) = 0$ otherwise, where $\hat{u}_{i^*} = \hat{u}_{i^*}(T) = \hat{d}$. Then, the effective dynamics look like $\hat{d}(t) = \text{exp}(\int_0^t ds \lambda_{i^*}(s)) \hat{d}(0)$. This results in an effective gain or decay on the system mode $\hat{d}$, depending on the sign of the corresponding eigenvalue integrated in time.

Numerically, we observed that filter optimization results in a bandpass filter with a bandwidth $B$, where the dissipation rate is $\kappa_{\text{eff}} \approx 0$ for $|\Delta_{\text{bs, tm}}| > \frac{B}{2}$, and $\kappa_{\text{eff}} \approx 2\pi g_{\text{bs, tm}}^2/B$ otherwise. More details about the filter optimization can be found in Appendix \ref{app:filter_optimization}. 

For the multi-level anharmonic oscillator case, we aim to simultaneously engineer gain or decay on multiple transitions with differing frequencies. For this purpose, we consider the analogous model of multiple Gaussian modes with frequencies $\omega_{d,1}, \omega_{d,2}, \dots$. We show in Appendix \ref{app:independent_dissipators} that if these modes are far detuned enough, i.e. $|\omega_{d,i} - \omega_{d,j}| \gg g_{\text{bs, tm}}^2/B$, the dissipation is independent for each mode. In other words, gain or decay is achieved for each mode independently using the same filter, with Lindblad dissipators $\kappa_\text{eff,i}\mathcal{D}[\hat{d}_i]$ ($\kappa_\text{eff,i}\mathcal{D}[\hat{d}_i^\dagger]$) for the decaying (gain-experiencing) modes, where $\kappa_\text{eff,i}$ is the effective dissipation rate stated above. This independence is what allows a single filter to realize the level-resolved dissipation structure assumed in Sec.~\ref{sec:0_n_qubit}.

\subsection{Implementation with the Kerr Hamiltonian}
\label{subsec:kerr_hamiltonian}

Building on the results from the previous subsection, showing how we can simultaneously engineer loss or gain on multiple modes with distinct frequencies, we now show how to implement the required dissipation for the 0-n Fock qubit.
To implement frequency-selective dissipation, we consider a realization based on the Kerr nonlinearity. The qubit Hamiltonian is \cite{Koch2007Transmon}
\begin{align}
    H = \omega_d \hat{a}^\dagger \hat{a} - K (\hat{a}^\dagger)^2\hat{a}^2
\end{align}
where $K$ is the self-Kerr nonlinearity. The energies of Fock levels are given by $E_n = \omega_d\, n - K n (n-1)$ for $n \geq 0$. Defining the transition frequency between adjacent levels as $f_{i(i+1)} \coloneqq E_{i+1} - E_i$, we find $f_{i(i+1)} = \omega_d - 2K i$ for $i \geq 0$. The fundamental transition frequency is $f_{01} = \omega_d$, and the anharmonicity is $\eta \coloneqq f_{12} - f_{01} = -2K$. To connect with transmon parameters in the regime $E_J/E_C \gg 1$, we identify $\omega_d \approx \sqrt{8E_J E_C}-E_C$ and $K \approx E_C/2$. With this parametrization, adjacent transitions are separated in frequency by $|f_{i(i+1)} - f_{(i+1)(i+2)}| = 2K \approx E_C$.

The key observation is that each transition $|i\rangle \leftrightarrow |i+1\rangle$ occurs at a distinct frequency due to the anharmonicity. This allows individual transitions to be addressed independently, provided the frequency separation exceeds the effective linewidth of the engineered dissipation. Specifically, if $E_C \gg g_{\text{bs, tm}}^2/B$, each transition can be treated as a separate mode coupled to the filter.

The detunings in Eq.~(\ref{eq:qubit_filter_hamiltonian}) for the transition between levels $|i\rangle$ and $|i+1\rangle$ are:
\begin{subequations}
\label{eq:detunings_defn}
\begin{align}
    \Delta_{\text{bs},i} &= \omega_f + \omega_{\text{bs}} - \omega_d + 2Ki, \\
    \Delta_{\text{tm},i} &= \omega_f + \omega_{\text{tm}} + \omega_d - 2Ki.
\end{align}
\end{subequations}
The detuning for the beam splitter coupling increases with Fock level index, while it decreases for the two-mode squeezing coupling. By choosing the drive frequencies $\omega_{\text{bs}}$ and $\omega_{\text{tm}}$ appropriately, we can place the transitions in the lower half of the qubit inside the filter passband (enabling engineered decay) while keeping the upper-half transitions outside the passband (and vice versa for gain). The placement of transitions relative to the filter passband is illustrated in Fig.~\ref{fig:filter_b}, where engineered decay is shown in red and gain in blue.

\subsection{Parameter scaling and constraints}
\label{subsec:parameter_scaling}

We now analyze the constraints on system parameters and how they scale with the number of levels $N$.

\paragraph{Transmon well depth.}
Implementing the 0-n Fock qubit with a transmon requires a sufficiently deep cosine potential to confine $N$ levels. The number of confined levels scales as $N_\text{levels} \sim \sqrt{E_J/(2E_C)}$ for $E_J/E_C \gg 1$~\cite{wang_high-_2025}. Fixing the fundamental transition frequency at $f_{01} = 5$~GHz together with the level-confinement requirement $E_J/(2E_C) \approx N^2$ uniquely determines $E_J$ and $E_C$ as functions of $N$.


\paragraph{Scaling with $N$.}
For $N \gg 1$, we have $f_{01} \approx 4 N E_C$, which implies $E_C \propto 1/N$. Since the bandwidth scales as $B \propto E_C N$, it approaches a constant value for large $N$. 
The independent dissipation assumption requires $E_C \gg g^2/B$, or $g^2 \ll E_CB \sim E_C^2 N$. Setting $g^2 \sim E_C^2 N$ (with appropriate prefactors ensuring $g^2/E_C^2 \ll N$),
the coupling strength decreases as $g^2 \propto 1/N$. Consequently, the effective engineered dissipation rate scales as
\begin{align}
    \kappa_{\text{eff}} \sim \frac{g^2}{B} \propto \frac{1}{N}.
\end{align}
Therefore, the stabilization time, i.e. the timescale $\sim 1/\kappa_{\text{eff}}$ over which the engineered dissipation returns population from unwanted intermediate levels to the code manifold $\{\ket{0},\ket{n}\}$, increases linearly with $N$. This timescale is relevant because it contributes to the duration of each syndrome-extraction cycle when the 0-n Fock qubit is used for error correction (see Sec.~\ref{sec:0_n_performance}).

In the regime dominated by intrinsic loss, plugging $\kappa_i^\downarrow/\kappa_s = \beta N$ into Eq.~\eqref{eq:bit_flip_rate_spectral_gap} and minimizing over $N$, the minimal 
bit-flip rate is achieved for
$N^* \approx 1/(e\beta)$. Beyond this value, increasing the number of levels $N$ is not beneficial for the suppression of bit-flips. Since our regime of interest is $\beta N \ll 1$, we operate well below this critical value. 
We also note that this scaling follows from fixing $f_{01}$, which forces
$E_C \propto 1/N$. If instead $E_C$ is held fixed while $E_J$ is increased with $N$
(as in Sec. \ref{subsec:logical_error_rates}, where this suppresses charge-noise dephasing), the bandwidth $B$ is
fixed and $g^2/B$ need not decrease, so $\kappa_{\rm eff}$ can be kept approximately
constant.

\paragraph{Filter roll-off and out-of-band suppression.}
Outside the filter bandwidth, residual engineered dissipation persists due to the finite roll-off of the filter response, as visible for $|\Delta| > B/2$ in Fig.~\ref{fig:filter_b}. The roll-off can be steepened by increasing the number of filter modes $M$ or by optimizing the couplings $\{J_i\}$ and bath dissipation $\kappa_f$ (see Appendix~\ref{app:filter_optimization}). For the parameters considered here, $M \sim 7$ modes suffice to suppress unwanted transitions below the intrinsic loss rates. We elaborate on this in Sec.~\ref{sec:0_n_performance}.

\paragraph{Protection against excitation above $|n\rangle$.}
The 0-n Fock qubit is susceptible to intrinsic heating, which can excite population to levels above $|n\rangle$. While our design does not actively protect against such processes, intrinsic relaxation from these higher levels back to $|n\rangle$ provides natural suppression. However, it is crucial that engineered gain to levels above $|n\rangle$ remains below the intrinsic decay rates. This is demonstrated in Fig.~\ref{fig:filter_b} for the $|5\rangle \to |6\rangle$ transition. Achieving sufficient suppression requires $M \sim 7$ filter modes for the parameters that we consider.

\section{0-n qubit as an ancilla}
\label{sec:0_n_performance}


Having established that the 0-n Fock qubit exhibits noise bias and can be realized with current experimental tools, we now consider its use as an ancilla for cat qubit architectures in order to obtain noise-biased gates. In such architectures, the ancilla mediates syndrome extraction by coupling to the data qubits through entangling gates.
Repetition codes are a natural choice for biased-noise qubits, as when bit-flip errors are exponentially suppressed, the dominant remaining errors are phase flips, which can be corrected using these codes \cite{Guillaud2019RepetitionCat, Gouzien2023RepetitionCatArchitecture, LeRegent2023RepetitionCat}. 
In such implementations, logical errors are corrected by measuring $X$ stabilizers, which require controlled-$X$ (CX) gates applied to data qubits, controlled by the ancilla.

$X$ errors on the ancilla during syndrome extraction using CX gates can propagate to the cat qubit, limiting the achievable noise bias \cite{Aliferis2008, puri_bias-preserving_2020}. The 0-n Fock qubit, as it is inherently noise biased, can surpass this limitation compared to architectures where the ancilla is a standard transmon qubit (without any engineered dissipation) \cite{hann_hybrid_2025}. Moreover, the simple, native dispersive coupling between the Fock qubit and the cat data qubits suffices to implement the CX gate. 

In light of this, we examine the performance of a CX gate conditioned on the 0-n Fock qubit, where we define a bit-flip in the following manner. Starting from the logical state $\ket{0_L}$ of two cat qubit resonators, and the $\ket{+_{0n}} = \frac{1}{\sqrt{2}}(\ket{0}+\ket{n})$ state of the ancilla, two CX gates are implemented to measure the $XX$ stabilizer of the resonators in the repetition code. The cat qubits are then stabilized to the code space via a two-photon dissipation process. 
A bit-flip occurs if the resulting state of the cat (code) qubits has odd parity, i.e. if they are in the $\ket{0_L} \otimes \ket{1_L}$, or $\ket{1_L} \otimes \ket{0_L}$ state.

Physically, the bit-flip probability arises because intermediate Fock populations ($0 < k < n$) contribute partial phase-space rotations to the cat qubit that do not map $\ket{0_L}$ cleanly onto itself or $\ket{1_L}$, leaving residual overlap with the wrong logical state after its stabilization. The role of the engineered dissipation is therefore to suppress these intermediate populations, concentrating the ancilla in the $\{\ket{0}, \ket{n}\}$ manifold, where the CX gate acts as either the identity or a full bit-flip. 

The use of the 0-n Fock qubit as an ancilla is demonstrated in Fig. \ref{fig:results_fig_a}. First, the superposition state $\ket{+_{0n}}$ is prepared using $X$ pulses between adjacent levels. The qubit is then stabilized dissipatively, and coupled to stabilized cat qubits. CX gates conditioned on the ancilla state are implemented sequentially, after which the cat qubits are stabilized, and the ancilla is measured in the $X$ basis.

Note that one can also consider a fully Fock-qubit architecture where both data qubits and ancilla are implemented accordingly. However, this approach would require bias-preserving gates between Fock qubits, including the logical X operation connecting $\ket{0}$ to $\ket{n}$. Such an operation is not straightforward to implement experimentally, thus, we focus here on only a Fock qubit ancilla. We discuss possible implementations of gates between Fock qubits in Appendix \ref{app:fock_qubit_gates}.

\begin{figure*}
    \centering
\includegraphics[width=0.95\linewidth]{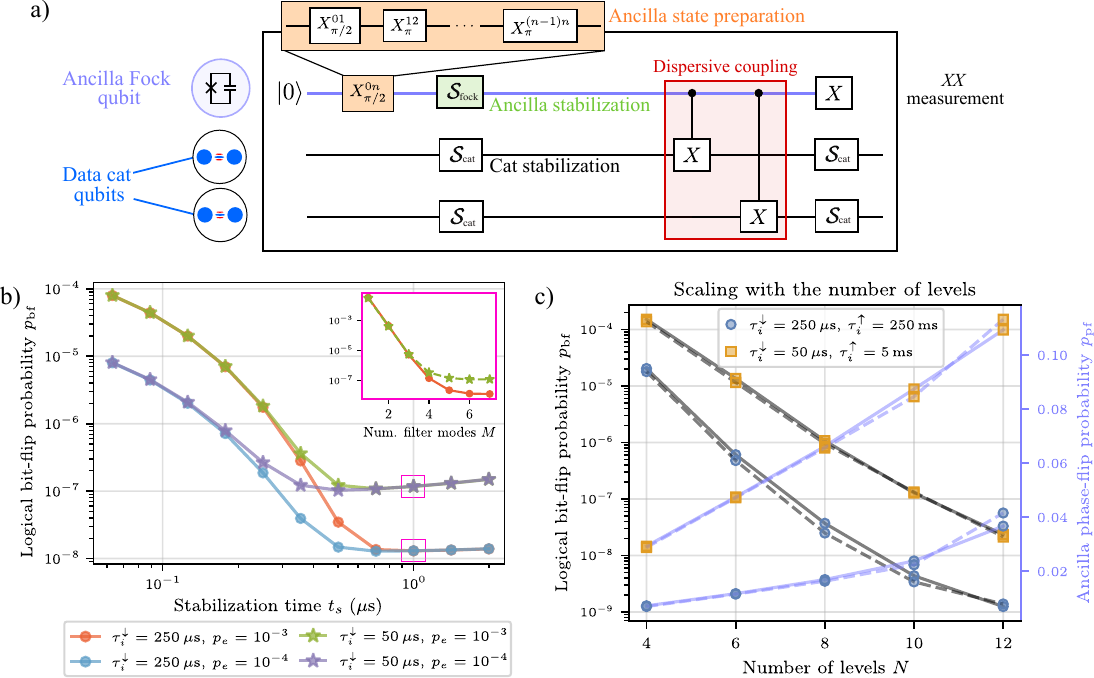}
    \caption{Use of the 0-n Fock qubit as an ancilla for noise-biased CX gates on stabilized cat qubits. a) Schematic of the protocol. The ancilla is prepared in the $\ket{+_{0n}} = \frac{1}{\sqrt{2}}(\ket{0}+\ket{n})$ state using $X$ pulses, then dissipatively stabilized (indicated by $\mathcal{S}_\text{fock}$) using the implementation in Sec. \ref{sec:filter_implementation}. Two CX gates are applied sequentially to two stabilized cat qubits (the stabilization is indicated by $\mathcal{S}_\text{cat}$) coupled dispersively to the ancilla, after which the cat qubits are restabilized and the ancilla is measured in the X basis. b) Logical bit-flip probability $p_\text{bf}$ of the CX gate as a function of the stabilization time $t_s$, for $n=7$, $|\alpha|^2 = 10$, and a filter with $M = 7$ filter modes. We assume perfect CX gates and consider two intrinsic noise configurations ($\tau_i^\downarrow = 50 \;\mu\text{s}, \, \tau_i^\uparrow = 5$ ms, and $\tau_i^\downarrow = 250 \;\mu\text{s}, \,\tau_i^\uparrow = 250 $ ms) and pulse error probabilities ($p_e = 10^{-4}$, $p_e = 10^{-3}$). $p_\text{bf}$ decays exponentially with stabilization time before converging to the bit-flip floor. \textit{Inset}: dependence of $p_\text{bf}$ on $M$, showing that $\sim 7$ modes suffice to suppress spurious engineered dissipation. c) $p_\text{bf}$ and ancilla phase-flip probability $p_\text{pf}$ as a function of the number of levels $N=n+1$ in the Fock qubit. We consider the intrinsic single-photon transitions during the CX gates. The gate time is assumed as $t_g = 200$ ns, $p_e = 10^{-3}$, and $|\alpha|^2 = 10$. The dashed lines correspond to analytic predictions. Bit-flip probabilities below $10^{-8}$ can be obtained with $N \geq 10$ levels. }
\label{fig:results_fig}
\sublabel{fig:results_fig_a}
\sublabel{fig:results_fig_b}
\sublabel{fig:results_fig_c}
\end{figure*}

\subsection{Bit-flip probability for the Fock-cat implementation}

Let us assume capacitive coupling between the 0-n Fock qubit and the cat qubit. If the coupling strength is weak relative to their detuning, i.e. $g \ll \Delta$, they can exhibit dispersive coupling in the form of $H = \chi_0 (\hat{a}^\dagger \hat{a}) \otimes (\hat{d}^\dagger \hat{d})$ (we set $\hbar=1$), where $\hat{a}$ and $\hat{d}$ are the annihilation operators of the Fock and cat qubit, respectively. The coupling strength is of order $\chi_0 \sim K g^2/\Delta^2$ \cite{Koch2007Transmon}. 
The evolution under this interaction will result in a rotation of the cat qubit in the phase space, proportional to the level number of the Fock qubit, in the form of:
\begin{align}
    e^{-i\chi_0 t (\hat{a}^\dagger \hat{a}) \otimes (\hat{d}^\dagger \hat{d})} \ket{m} \otimes \ket{\alpha}
    &=\ket{m} \otimes \ket{\alpha e^{-i \chi_0 m t}}.
\end{align}
For $m=0$, the cat qubit is unaffected, so the gate acts as the identity. In contrast, if $m\chi_0 t = \pi$, $\ket{\alpha} \leftrightarrow \ket{-\alpha}$, hence $\ket{0_L} \leftrightarrow \ket{1_L}$ and this operation implements the $X$ operator. Conditioning the cat rotation on the Fock-qubit level in this way realizes a CX gate controlled by the Fock qubit.

We define the logical bit-flip probability $\tilde{p}_\text{bf}(\theta)$ as a function of the rotation angle $0 \leq \theta \leq \pi$:
\begin{subequations}
\label{eq:bit_flip_probability_defn}
\begin{align}
    \tilde{p}_\text{bf}(\theta) &= \frac{1}{2} \left[ 1 - \left(1- M_\alpha(\theta)  \right)^2 \right], \\
    M_\alpha(\theta) &\coloneqq  \frac{2|\alpha|^2}{\sinh{(2|\alpha|^2)}} \int_{0}^{\theta} d\phi \, \sin{(\phi)} I_0\left(2|\alpha|^2\sin{(\phi)} \right)
\end{align}
\end{subequations}
For more details, see Appendix \ref{app:bit_flip_calcs}.
Furthermore, for a superposition over different Fock states of the control qubit, the total bit-flip probability is a linear function of the bit-flip probability for each Fock state, as coherence terms in the form of $\ket{m}\bra{n}, m \neq n$, will not contribute when the control qubit is traced out. Since the rotation angle in phase space is proportional to $m$ for the $m^\text{th}$ Fock level,
if we set the rotation angle to be $\theta=\pi$ for the highest level $\ket{n}$,
the total bit-flip probability $p_\text{bf}$ is found as
\begin{align}
\label{eq:total_bit_flip_prob}
    p_\text{bf} = \sum_{k=0}^n \tilde{p}_\text{bf}\left(\frac{\pi k}{n}\right) P(k)
\end{align}
where $P(k) \coloneqq \bra{k}\hat{\rho}\ket{k}$ is the population at the $k^\text{th}$ level. 
We show in the following subsections that without stabilization, increasing n populates more intermediate levels through pulse errors, each of which contributes a partial rotation that degrades the cat qubit. Engineered dissipation, as described in Sec. \ref{sec:filter_implementation}, resolves this by pumping intermediate populations back to $\ket{0}$ and $\ket{n}$ before the CX gate is applied. 
Accordingly, we first examine the bit-flip probability of the 0-n Fock qubit in the absence of engineered dissipation in Subsec. \ref{subsec:bf_prob_before_stabilization}. Then, we present the numerical results with stabilization in Subsec. \ref{subsec:bf_prob_after_stabilization}.

\subsubsection{Bit-flip probability without stabilization}
\label{subsec:bf_prob_before_stabilization}

Our aim is to prepare the superposition state $(\ket{0}+\ket{n})/\sqrt{2}$ with imperfect pulses. Let us define $\ket{\pm_{ab}} \coloneqq (\ket{a} \pm \ket{b})/\sqrt{2}$.
We first prepare the superposition state $\ket{\pm_{01}}$ from the ground state $\ket{0}$ by performing an imperfect $\pi/2$ pulse. We assume the following Kraus operators during this operation:
\begin{subequations}
\begin{align}
    \hat{K}_0 &= \sqrt{1-p_e} \, \left( \identity-\identity_{01}+ \ket{+_{01}}\bra{0} + \ket{-_{01}}\bra{1} \right) \\ \hat{K}_1 &= \sqrt{p_e} \, \identity
\end{align}
\end{subequations}
where $\identity \coloneqq \sum_m \ket{m}\bra{m}$ is the identity operator across all levels, and $\identity_{01} = \ket{0}\bra{0} + \ket{1}\bra{1}$ is the identity operator for the $\{\ket{0},\ket{1}\}$ subspace. Then, starting from the initial state $\ket{0}\bra{0}$, the resulting qubit state will be $(1-p_e)\ket{+_{01}} \bra{+_{01}} + p_e \ket{0}\bra{0}$. 
The remaining pulses will be $\pi$ pulses performed between levels $\ket{m}$ and $\ket{m+1}$, for $1\leq m < n$. We assume the following Kraus operators:
\begin{subequations}
\begin{align}
    \hat{K}_0^{m} &= \sqrt{1-p_e} (\identity - 2\ket{-_{m(m+1)}}\bra{-_{m(m+1)}}),  \\
    \hat{K}_1^{m} &= \sqrt{p_e} \, \identity
\end{align}
\end{subequations}
After the $\pi$ pulses, the state of the control qubit can be written as
\begin{align}
\label{eq:control_qubit_imperfect_pulse_final_state}
    \hat{\rho}_n &= (1-p_e)^n \ket{+_{0n}} \bra{+_{0n}} + p_e \ket{0} \bra{0} \nonumber \\
    &\quad + p_e \sum_{i=1}^{n-1} (1-p_e)^{i} \ket{+_{0i}} \bra{+_{0i}} 
\end{align}
Using this state as the control, we can perform CX gates on two resonators encoded in the form of cat qubits. After this gate, the total bit-flip probability is found as (see Appendix \ref{app:bit_flip_calcs})
\begin{align}
    p_\text{bf} &\approx \frac{p_en}{\sqrt{2 \pi^3}|\alpha|}
\end{align}
for $|\alpha|, n \gg 1$. Then, the bit-flip probability of the CX gate grows linearly with $n$ and $p_e$ up to first order in $p_e$: this scaling demonstrates the need for stabilization on the 0-n Fock qubit in order to obtain suppression of bit-flip probabilities with larger $n$.

\subsubsection{Bit-flip probability with stabilization}
\label{subsec:bf_prob_after_stabilization}

Engineered dissipation ensures the stabilization of the 0-n Fock qubit to minimize the bit-flip probability. First, notice that since the bit-flip probability is independent of the coherences of the 0-n Fock qubit, we can work with the CTMC description to compute the steady state of the qubit in the idealized case (cf. Sec. \ref{sec:0_n_qubit}). For short timescales $t \sim 1/\kappa_s$, the spectral gap is negligible if $\kappa_i^\downarrow, \kappa_i^\uparrow \ll \kappa_s$, and the qubit converges to the metastable states for the upper and lower sections independently. 
Then, the bit-flip probability also converges to a fixed value determined by the population distribution across the qubit: we refer to this value as the \textit{bit-flip floor}. In this analysis, we disregard the heating to levels above $\ket{n}$, which is negligible for sufficiently short stabilization times.

The bit-flip floor depends on the highest level number $n$, mean photon number of the cat qubits $|\alpha|^2$, and the relative strength of intrinsic loss processes $\kappa_i^\downarrow/\kappa_s$, $\kappa_i^\uparrow/\kappa_s$. We observe that this floor decays exponentially with $n$ for relatively small $n$, whereas the scaling is modified to $1/n^2$ in the limit of $n \gg 1$. For a fixed $n$, the bit-flip probability decays exponentially with stabilization time $t_s$ before converging to the bit-flip floor. For an extended discussion on the bit-flip floor, see Appendix \ref{app:bit_flip_floor}.

We analyze the convergence to the bit-flip floor numerically in Fig. \ref{fig:results_fig_b}. We fix $n=7$, $|\alpha|^2 = 10$, and examine the logical bit-flip probability $p_\text{bf}$ as a function of the stabilization time, $t_s$, for two noise configurations. These configurations display distinct intrinsic decay and heating times $\tau_i^{\downarrow, \uparrow} = 1/\kappa_i^{\downarrow,\uparrow}$. For the first one, we assume $\tau_i^\downarrow = 50 \;\mu$s and $\tau_i^\uparrow = 5 $ ms, whereas for the second one, $\tau_i^\downarrow = 250 \;\mu$s and $\tau_i^\uparrow = 250 $ ms. Furthermore, we consider pulse error probabilities of $p_e = 10^{-3}$ and $p_e = 10^{-4}$. 
Lastly, we use a filter with $M = 7$ modes. For long stabilization times where the bit-flip probability approaches the bit-flip floor (e.g. $t_s = 1 \; \mu$s), we observe that this number is sufficient to suppress spurious engineered dissipation between levels, as can be seen in the inset of this Figure. For these times, bit-flip probabilities are independent of the pulse error probability $p_e$, in contrast to the case at the initialization (i.e. $t_s \ll  1 \, \mu$s). For longer stabilization times, the bit-flip probabilities start increasing due to spurious heating to levels above $\ket{n}$, which are not inhibited via the engineered dissipation. We note that this Figure assumes a perfect CX gate with no intrinsic decay or heating events.

In Fig. \ref{fig:results_fig_c}, we plot the logical bit-flip probability $p_\text{bf}$, as well as the ancilla phase-flip probability $p_\text{pf}$, as a function of the number of levels $N$ in the Fock qubit. For the latter, the cat qubits are initialized in state $\ket{+_L} = (\ket{0_L} + \ket{1_L})/\sqrt{2}$, and the Fock qubit is initialized as $\ket{+_{0n}}$. After subsequent CX gates conditioned on it, the Fock qubit is measured in the $X$ basis, and the phase-flip probability is defined as $p_\text{pf} = 1-\bra{+_{0n}}\hat{\rho}\ket{+_{0n}}$, where $\hat{\rho}$ is the final state of the qubit. 
In Eq. (\ref{eq:phase_flip_prob}), we have derived that for the idealized case of single-photon transitions between adjacent levels due to loss or engineered dissipation, the phase-flip probability grows linearly with $n$ for $\tilde\kappa  t \ll 1$, where $\tilde \kappa=n \kappa_i^\downarrow + \kappa_i^\uparrow \approx n \kappa_i^\downarrow$
is the total depopulation rate of level $\ket{n}$ (assuming $\kappa_i^\uparrow \sim \kappa_i^\downarrow$ and $n \gg 1$), and $t$ is the total time dedicated to stabilization and the CX gates. This property is in line with other noise biased qubits, where bit-flips are suppressed exponentially with $n$, whereas phase-flips are enhanced linearly with $n$.

We now quantify how the bit-flip and phase-flip probabilities scale with the number of levels $N$ when intrinsic single-photon events during the CX gate are included (referred to as the imperfect CX gate), showing that bit-flips remain exponentially suppressed, approaching $10^{-8}$ for $N\geq10$, while phase-flips grow only linearly. 
We set the per-pulse error as $p_e = 10^{-3}$, mean photon number of cat qubits as $|\alpha|^2 = 10$, the Fock qubit stabilization time as $t_s = 1 \, \mu$s, and consider the same noise configurations as above. The two configurations are indicated with yellow square markers and blue circular markers, for high and low noise, respectively. 
We assume a CX gate time of $t_g = 200$ ns, which is experimentally feasible \cite{puri_bias-preserving_2020, hann_hybrid_2025, putterman_hardware-efficient_2025}. For both the logical bit-flips and the ancilla phase-flips, we plot the analytically calculated probabilities with dashed lines of the respective color. For this purpose, we use the approximate engineered dissipation rates obtained perturbatively (cf. Appendix \ref{app:filter_optimization}), and use the CTMC description of the evolution of the diagonal elements of the Fock qubit. We show in Appendix \ref{app:analytical_CX} that the coherences of this qubit do not affect the bit-flip probability when the evolution can be described with the single-photon jump operators, as in Eq. (\ref{eq:lindblad_eqn}). For this setting, we propose a solution for the evolution of the bit-flip probability which also takes into account single-photon events during the gate. We observe that the analytical prediction matches the numerical results for all cases. Thus, the assumptions of (i) independent dissipators for each transition within the Fock qubit, and (ii) adiabatic evolution of the bare system mode hold in this regime.

In summary, taking into consideration all intrinsic single-photon events during the ancilla stabilization and the CX gate, bit-flip probabilities below $10^{-8}$ can be achieved with a Fock qubit with $N \geq 10$ levels. The bit-flip probability is approximately exponentially suppressed with $N$, mirroring the scaling of cat qubits with mean photon number $|\alpha|^2$, achieved through a complementary mechanism: the separation between levels $\ket{0}$ and $\ket{n}$ in a multi-level, anharmonic oscillator, rather than the separation of coherent states in phase space. 
These results indicate that the 0-n Fock qubit ancilla can provide exponential bit-flip suppression comparable to cat-cat architectures, 
while only requiring engineered beam splitter and two-mode squeezing couplings for engineered dissipation. Moreover, simple, native dispersive couplings suffice to implement CX gates between Fock and cat qubits that preserve both qubits' exponential noise bias.
This approach thus substantially simplifies the experimental requirements relative to bias-preserving cat-cat gates.

\subsection{Logical error rates with the repetition code}
\label{subsec:logical_error_rates}

\begin{figure}
    \centering
\includegraphics[width=\linewidth]{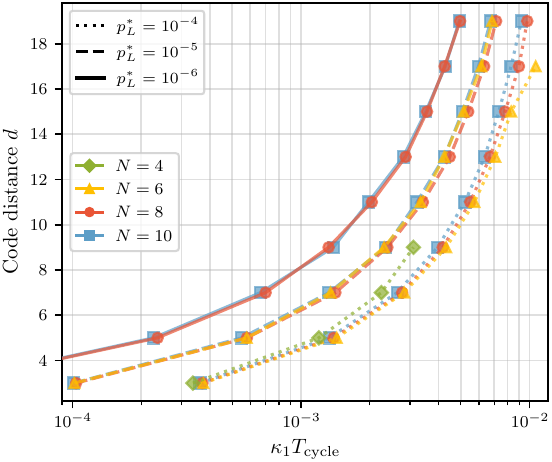}
    \caption{Average logical error per round $p_L^*$ when the 0-n Fock qubit is used as an ancilla in the repetition code, paired with cat data qubits.
    The contours for $p_L^* \in \{10^{-4}, 10^{-5}, 10^{-6}\}$ are shown as a function of $\kappa_1 T_\text{cycle}$ and the code distance $d$, where
    $\kappa_1$ is the single-photon loss rate of the cat qubits, and $T_\text{cycle}$ is the cycle time for a single round of error correction. We work with a phenomenological noise model, where  $Z$ errors are applied on the ancilla and the data qubits at the start of each error correction round. The ancilla $Z$ errors occur only due to single-photon events ($\tau_i^\downarrow = 250 \;\mu\text{s}, \,\tau_i^\uparrow = 250 $ ms) during the stabilization ($t_s = 500$ ns), or the CX gates ($t_g = 200$ ns): we assume large enough $E_J/E_C$ such that dephasing due to charge noise is sufficiently suppressed.
    Lastly, the cat qubits have a mean photon number of $|\alpha|^2=10$. }
    \label{fig:rep_code_threshold}
\end{figure}

We now examine the logical error rates achievable when the 0-n Fock ancilla is used for syndrome extraction in a repetition code. Here, as in the previous subsection, syndrome measurements are performed using two CX gates conditioned on the ancilla applied to the neighboring data qubits, followed by measurement of the ancilla in the
X basis.

We simulate the performance of a distance-$d$ repetition code \cite{chamberland_building_2022, putterman_hardware-efficient_2025} under a phenomenological noise model in which both ancilla and data qubits undergo $Z$ errors at the start of each round, prior to the CX gates \cite{Dennis2002TopologicalMemory}. For the cat data qubits, the dominant mechanism is single-photon loss, which induces $Z$ errors with probability $p_Z^\text{data} = (1-e^{-2|\alpha|^2\kappa_1 T_\text{cycle} )}/2 \approx |\alpha|^2 \kappa_1T_\text{cycle}$
in the regime $|\alpha|^2 \kappa_1T_\text{cycle} \ll 1$, where $\kappa_1$ is the loss rate, and $T_\text{cycle}$ is the duration of one syndrome extraction round. Similarly, $Z$ errors on ancilla qubits can occur due to single-photon loss and heating events, or dephasing. The latter can be mitigated by operating in the large $E_J/E_C$ regime, where $E_J$ is increased and $f_{01}$ is adjusted so that $E_C$ remains fixed, exponentially suppressing charge-noise-induced dephasing \cite{wang_high-_2025}. Therefore, we ignore dephasing effects and consider only intrinsic single-photon events. A more pessimistic noise budget that includes dephasing is presented in Appendix \ref{app:logical_error_with_dephasing}. 

In order to combat the effect of single-photon events as much as possible, we shorten the stabilization time to $t_s = 500$ ns, which is sufficient to approach the bit-flip floor (see Fig. \ref{fig:results_fig_b}). The gate time is set to $t_g = 200$ ns as in the previous subsection, and the ancilla preparation (before the stabilization) is still subject to pulse errors with probability $p_e = 10^{-3}$. The ancilla $Z$ error probability per round is thus set entirely by $t_s$, $t_g$, and $p_e$, and does not vary with $T_\text{cycle}$: only the data-qubit error probability $p_Z^\text{data}$ scales with $T_\text{cycle}$. Note that the ancilla preparation time scales with the highest level number $n$, as the pulses are assumed to be sequential. Each selective pulse is assumed to take $\sim 50$ ns.
Furthermore, syndrome measurements are assumed to be subject to a readout error with probability $p_\text{meas} = 0.01$, and we consider the lower intrinsic noise configuration from the previous subsection, with $\tau_i^\downarrow = 250 \;\mu$s and $\tau_i^\uparrow = 250 $ ms. Lastly, the mean photon number of a cat qubit is taken as $|\alpha|^2 = 10$.

Given these parameters, we simulate $d$ rounds of a distance-$d$ repetition code, and find the logical error rate per round occurring due to phase-flip events. We then  compute the overall logical error rate $p_L^*$ by adding the residual bit-flip contribution per round, which is found assuming imperfect CX gates with intrinsic single-photon events. The bit-flip contribution is found by taking the per-stabilizer bit-flip probability, as found in Fig. \ref{fig:results_fig_c} in the previous subsection, and multiplying it by the number of stabilizer measurements per round, $d-1$.
In Figure \ref{fig:rep_code_threshold}, we plot the contours along which the overall logical error rate reaches $10^{-4}$, $10^{-5}$ and $10^{-6}$ as a function of $\kappa_1 T_\text{cycle}$ and the code distance $d$, for $N \in \{4,6,8,10\}$. The various logical error rate contours are shown with dotted, dashed, and plain lines, whereas the various $N$ are plotted with different colors. From the Figure, we observe that for a given logical error rate, the contours for different $N$ overlap significantly, which signifies that data phase-flips are the dominant source of error in this regime, rather than ancilla phase-flips. 

Thus, thanks to the exponential suppression of bit-flips achieved by the 0-n Fock ancilla, we can achieve error rates as small as $10^{-6}$. Note that this target is achieved for $N\geq 8$ due to the bit-flip contribution to the logical error surpassing the target value for smaller $N$. In contrast, we are able to reach higher logical error rates with these $N$, for example, $p_L^*=10^{-4}$. For comparison, achieving $p_L^*=10^{-6}$ with a standard transmon ancilla, which lacks exponential noise bias, would require the ability to correct ancilla-transmon-induced bit flips with a more hardware-intensive code, like the surface code.

The $p_L^* = 10^{-6}$ logical error rate (i.e. the megaquop regime \cite{Preskill_megaquop}) is achieved in the experimentally realistic regime of $\kappa_1 T_\text{cycle} \sim 10^{-3}$, corresponding to a storage-mode lifetime of $1/\kappa_1 \sim$2 ms for the $T_\text{cycle} \sim 2 \, \mu$s cycle times considered here. Single-photon lifetimes exceeding 10 ms have been demonstrated in bare superconducting cavities \cite{Reagor2013, Milul2023}, and current cat-qubit experiments operate within an order of magnitude of this target \cite{putterman_hardware-efficient_2025}. Furthermore, if one allows for a larger code distance of $d \sim 17$, this target can be achieved with $\kappa_1 T_\text{cycle} \sim 4\cdot10^{-3}$, corresponding to a lifetime of $1/\kappa_1 \sim 500 \, \mu$s.

The cycle time can in principle be reduced significantly further by pipelining the ancilla operations across consecutive rounds. While one ancilla is engaged in the CX gates and subsequent syndrome extraction, the next ancilla can be prepared and stabilized in parallel, so that ancilla preparation no longer contributes to the cycle time. Moreover, because each ancilla is decoupled from the filter once its stabilization is complete, the same multimode filter can be shared between consecutive ancillas rather than requiring a dedicated filter per ancilla. Note that this approach requires two sets of ancillas that are used sequentially. This pipelining could halve the effective cycle time to $\sim 1 \,\mu$s without modifications to the underlying hardware, bringing the operating regime to $\kappa_1 T_\text{cycle} \sim 10^{-3}$ for $1/\kappa_1 \sim 1$ ms. An alternative approach to shorten the cycle time, in principle, is to state transfer the Fock qubit ancilla to a separate element for performing the readout, so that it can be re-prepared for the next error correction cycle during the stabilization of the cat qubits \footnote{See e.g. Ref. \cite{chamberland_building_2022}, where the ancilla is transferred to a photonic readout mode and repeatedly measured via a quantum non-demolition (QND) readout to mitigate transmon errors.}.

These results demonstrate that the 0-n Fock ancilla preserves the noise bias of the cat data qubits well enough that ancilla-induced bit flips no longer dominate the logical error budget for moderate $N$. The hybrid cat-Fock architecture therefore provides a viable route to experimental realization of low logical error rates.

\section{Conclusions}

We have introduced the 0-n Fock qubit, a noise-biased bosonic encoding in which the logical states are defined by the Fock levels ${\ket{0},\ket{n}}$. The protection mechanism relies on the fact that single-photon loss events map the logical manifold to intermediate states rather than directly inducing bit flips. As a consequence, logical bit-flip errors require $\lceil n/2 \rceil$ sequential loss events and are therefore exponentially suppressed with increasing $n$, establishing a hardware-level noise bias analogous to the bit-flip suppression in cat qubits. However, distinct from cat qubits, the Fock qubit can be implemented with multi-level anharmonic oscillators, such as a transmon, or a Kerr nonlinear oscillator.

We have shown that this qubit can be stabilized using frequency-selective dissipation. By coupling the oscillator to a short chain of LC resonators, the effective bath spectral density can be shaped such that unwanted intermediate Fock states are selectively depopulated while the logical manifold remains invariant. Importantly, this stabilization can be realized with a small number ($<10$) of filter modes and realistic circuit parameters, indicating that the required bath engineering is compatible with current superconducting circuit technology. The gain-engineering mechanism underlying our stabilization has recently been demonstrated experimentally in a multilevel transmon \cite{gain_engineered_transmon}. There, a two-mode squeezing interaction between the transmon and a lossy resonator realizes a frequency-selective gain channel on the $\ket{e} \leftrightarrow \ket{f}$ transition that autonomously corrects single-photon loss from level $\ket{f}$, 
providing a viable $\ket{g}-\ket{f}$ encoding. Extending such a scheme to larger $n$ requires frequency-selective dissipation across multiple levels of the transmon, as realized here by the multimode filter.

Beyond its intrinsic protection properties, the 0-n qubit is particularly well suited for hybrid architectures with cat qubits. In dissipatively stabilized cat qubits, operations associated with the logical 
$X$ degree of freedom are naturally implemented through phase-space rotations, whereas the logical $Z$ operator requires distinguishing $\ket{\pm\alpha}$, which is experimentally more demanding.
In contrast, for the Fock qubit, the $Z$ operator
can be implemented straightforwardly via dispersive coupling to a resonator, while direct access to the logical $X$ degree of freedom is more difficult. Thus, a hybrid architecture combining cat data qubits with 0-n ancilla qubits enables efficient and bias-preserving CX gates, which are central to syndrome extraction in fault-tolerant protocols.

Accordingly, we studied the use of the 0-n qubit in CX gates for syndrome measurements on cat qubits, and found that bit-flip probabilities below $10^{-8}$ are achievable with a Fock qubit with $N=10$ levels under experimentally realistic conditions. We also simulated a repetition code using these gates, Fock ancilla qubits and cat data qubits, where we obtained logical error rates as low as $p_L^*=10^{-6}$ at a code distance of $d=9$ under similar conditions.
This demonstrates that the 0-n Fock qubit can serve as a bias-preserving ancillary resource without requiring the same degree of Hamiltonian engineering as fully bosonic cat-cat interactions, while avoiding the bias limitations associated with transmon-based hybrid architectures.

\begin{acknowledgments}
We acknowledge useful discussions with Ron Belyansky, Yoni Schattner, Ian Yang Yen Wei, Francesco Adinolfi, and
 Alexander Grimm.
The majority of the work was completed as part of S.D.'s internship at the Amazon Center for Quantum Computing, and the entirety of it was completed while K.N. was a researcher there.
\end{acknowledgments}

\appendix

\section{Filter optimization}
\label{app:filter_optimization}

From the Gaussian approximation, the effective decay or gain rate on the dressed mode that corresponds to the bare system mode is found perturbatively as
\begin{align}
\label{eq:gaussian_approximate_rate}
    \kappa_{\text{eff}}(\Delta) \approx 2g^2\left| \text{Im} \left( \; \Delta - \frac{J_{1,2}^2}{\Delta-\frac{J_{2,3}^2}{ \substack{\ddots\\\;\;\Delta - i\frac{\kappa_f}{2}} }} \right)^{-1}  \right|
\end{align}
given that $g \ll J_{i,i+1} \; \forall \,i$. Here, $g$ is the interaction rate with the filter ($g_{\text{bs},\text{tm}}$ in the main text), and $\Delta$ is the detuning ($\Delta_{\text{bs},\text{tm}}$ in the main text). For multiple modes, we can also show that the rates can be found this way using the respective interaction rates and detunings if the detuning of two modes $\Delta_i$ and $\Delta_j$ satisfy $|\Delta_i - \Delta_j| \gg \frac{g^2}{B}$. To optimize over the effective rate given $\{J_{i,i+1}\}$ and $\kappa_f$, we define the following filter cost function:
\begin{align}
    \mathcal{C}(r) &= \int_{-1/2}^{1/2} d\left(\frac{\Delta}{B}\right) \, \left(\frac{\kappa_{\text{eff}}(\Delta/B)}{\kappa_{\text{eff}}(0)} - 1 \right)^2 \nonumber \\
    &\quad + 2 r \int_{1/2}^{\infty} d\left(\frac{\Delta}{B}\right) \, \left(\frac{\kappa_{\text{eff}}(\Delta/B)}{\kappa_{\text{eff}}(0)} \right)^2.
\end{align}
This cost function is parametrized by $r$, which represents the ratio of importance of out-of-passband errors to in-passband errors. The roll-off speed of the filter increases with larger $r$, however, we observe larger-amplitude oscillations in $\kappa_\text{eff}$ in the passband. Then, $r$ can be tuned to change the ``flatness" of the filter. In the main text, we set $r\in [10^2, 10^3]$ while optimizing over the filter, for which we observed a satisfactory roll-off speed.

\section{Independence of dissipators}
\label{app:independent_dissipators}

In this Appendix, we show that the engineered dissipation induced by the multimode filter can be decomposed into independent Lindblad dissipators acting on each transition $\ket{k}\leftrightarrow\ket{k- 1}$ of the system mode, provided that the transition frequencies are sufficiently separated. We present two complementary derivations. In Appendix~\ref{app_sub:gaussian}, we treat the adjacent transitions as independent Gaussian modes coupled to the filter, and derive the independence condition $|\Delta_i - \Delta_j| \gg g^2/B$ via Heisenberg-picture perturbation theory. In Appendix~\ref{app_sub:non_gaussian}, we work directly with the anharmonic system Hamiltonian and adiabatically eliminate the filter modes using the effective operator formalism of Refs. \cite{reiter_effective_2012, chamberland_building_2022}. Both approaches yield the same independence condition and the same leading-order form for the effective dissipation rates.

\subsection{With the Gaussian approximation}
\label{app_sub:gaussian}

In order to simplify the dynamics analytically, we can approximate the transitions within the multilevel qubit as independent Gaussian modes with annihilation operators $\hat{d}_i$, $i=1,2,\dots$. We refer to these modes as the system modes. Similar to Eq. (\ref{eq:qubit_filter_hamiltonian}), the interaction Hamiltonian with respect to the bare Hamiltonian of the system modes, within the RWA, can be written as
\begin{align}
    \hat{H} &= g \hat{f}_1^\dagger \left( \sum_{j \in s_d}\hat{d}_j  \, e^{i \Delta_j t}  + \sum_{j \in s_g}\hat{d}_j^\dagger \, e^{i \Delta_j t} \right) 
    \nonumber \\ & \quad + \sum_{i=1}^{M-1} J_{i,i+1} \, \hat{f}_i \hat{f}_{i+1}^\dagger  + \text{h.c.}
\end{align}
where we assumed the same interaction rate $g$ for both the beam splitter and the two-mode squeezing interaction, $\{ \hat{f}_i\}$ are the filter annihilation operators, and $s_d, s_g$ denote the sets for which the respective modes experience engineered decay or gain, respectively. We also engineer dissipation on the final filter mode, with a dissipator $\kappa_f \mathcal{D}[\hat{f}_M]$. The evolution of the annihilation or creation operators under this Hamiltonian can be described within the Heisenberg picture: 
\begin{align}
    \frac{d}{dt} \begin{bmatrix}
        \hat{\mathbf{s}}(t) \\
        \hat{\mathbf{f}}(t)
    \end{bmatrix} &= \left(\mathbf{M}^{(0)} + g\mathbf{M}^{(1)} \right) \cdot \begin{bmatrix}
        \hat{\mathbf{s}}(t) \\
        \hat{\mathbf{f}}(t)
    \end{bmatrix} \nonumber \\
    &= B\left(\begin{bmatrix}
        \mathbf{S} &  \\
         & \mathbf{F}
    \end{bmatrix} + \tilde g \begin{bmatrix}
        \mathbf{0} & \mathbf{C}_1  \\
        \mathbf{C}_2 & \mathbf{0}
    \end{bmatrix} \right) \cdot \begin{bmatrix}
        \hat{\mathbf{s}}(t) \\
        \hat{\mathbf{f}}(t)
    \end{bmatrix}.
\end{align}
Here, $\hat{\mathbf{f}} \coloneqq (\hat{f}_1, \hat{f}_2, \dots,\hat{f}_{M-1})^T$ is the vector containing the annihilation operators of the filter modes, whereas $\hat{\mathbf{s}}$ is the vector containing the annihilation (creation) operators of the system modes in $s_d$ ($s_g$). With this convention, we can represent both dynamics in a single equation. Lastly, $B>0$ is the bandwidth of the filter, with $g \ll B$, $g \ll J_i \; \forall \, i$, and $g \ll \kappa_f$. We use the bandwidth to define $\tilde g \coloneqq g/B$ as a small, dimensionless perturbation.
 
We can choose a rotating frame such that $\mathbf{S}$ is a diagonal matrix containing the elements of $\{ \Delta_j \}$. 
$\mathbf{C}_1$ and $\mathbf{C}_2$ are the $(S, M)$ dimensional coupling matrices with $S = |s_d|+|s_g|$. For example, if all of the modes are coupled with the beam splitter (two-mode squeezing) interaction, $\mathbf{C}_1 = \mathbf{C}_2^T$ ($\mathbf{C}_1 = -\mathbf{C}_2^T$). Since the modes only couple to the first filter mode, only the first column of $\mathbf{C}_1$ and row of $\mathbf{C}_2$ are nonzero. The type of interaction determines the sign of the elements of the first column of $\mathbf{C}_1$, whereas the elements in the first row of $\mathbf{C}_2$ are always given by $-\tilde g$.  Lastly, due to the symmetry of $\mathbf{S}$ and $\mathbf{F}$, $\mathbf{M}$ is symmetric. However, $\mathbf{F}$ is a complex matrix if $\kappa_f > 0$: thus, the eigenvalues of $\mathbf{F}$ are complex in general.

We aim to quantify the effect of the coupling with the filter on the system modes. Notice that if all of the detunings $\Delta_j$ are distinct, the eigenvector matrix of $\mathbf{S}$ is the identity matrix, and the respective eigenvalues are $\Delta_j/B$. 
Let us write the perturbed, diagonal eigenvalue matrix as $\boldsymbol\Lambda^{(0)} + \tilde g \boldsymbol\Lambda^{(1)} + \tilde g^2 \boldsymbol\Lambda^{(2)}$, and the corresponding perturbed eigenvector matrix as $\mathbf{V}^{(0)} + \tilde g\mathbf{V}^{(1)} + \tilde g^2\mathbf{V}^{(2)}$. We assume that eigenvectors are the column vectors.
We also define submatrices in the form of
\begin{align}
    \boldsymbol\Lambda^{(i)} \coloneqq \begin{bmatrix}
        \boldsymbol\Lambda_S^{(i)} & \\
        & \boldsymbol\Lambda_F^{(i)}
    \end{bmatrix}, \quad \mathbf{V}^{(i)} \coloneqq \begin{bmatrix}
        \mathbf{V}_{SS}^{(i)} & \mathbf{V}_{SF}^{(i)} \\[1mm]
        \mathbf{V}_{FS}^{(i)} & \mathbf{V}_{FF}^{(i)}
    \end{bmatrix}.
\end{align}
Note that $\mathbf{V}_{SF}^{(0)} = \mathbf{V}_{FS}^{(0)} = \mathbf{0}$.
From grouping the terms on the same order in the eigenvalue equation, we first write
\begin{align}
    \mathbf{M}^{(0)} \mathbf{V}^{(1)} +  \mathbf{M}^{(1)} \mathbf{V}^{(0)} = \mathbf{V}^{(0)} \boldsymbol{\Lambda}^{(1)} + \mathbf{V}^{(1)} \boldsymbol{\Lambda}^{(0)} 
\end{align}
Multiplying from the left by $(\mathbf{V}^{(0)})^T$, we obtain
\begin{align}
    \boldsymbol\Lambda^{(1)} = \text{diag}((\mathbf{V}^{(0)})^T \mathbf{M}^{(1)} \mathbf{V}^{(0)}) = \mathbf{0},
\end{align}
and
\begin{align}
    &\mathbf{M}^{(0)} \mathbf{V}^{(1)} - \mathbf{V}^{(1)} \boldsymbol{\Lambda}^{(0)} = - \mathbf{M}^{(1)} \mathbf{V}^{(0)}. 
\end{align}
Then, we notice that on the first order, the eigenvalues are not modified due to the connection with the filter modes. For the first order perturbation in the eigenvectors, we find
\begin{subequations}
\label{eq:off_diag_perturb}
\begin{align}
\mathbf{S} \mathbf{V}_{SS}^{(1)} -   \mathbf{V}_{SS}^{(1)} \boldsymbol{\Lambda}_S^{(0)} &= \mathbf{0} \\
\mathbf{F} \mathbf{V}_{FF}^{(1)} -   \mathbf{V}_{FF}^{(1)} \boldsymbol{\Lambda}_F^{(0)} &=   \mathbf{0} \\
\mathbf{F} \mathbf{V}_{FS}^{(1)} -   \mathbf{V}_{FS}^{(1)} \boldsymbol{\Lambda}_S^{(0)} &= -\mathbf{C}_2 \mathbf{V}_{SS}^{(0)} \\
\mathbf{S} \mathbf{V}_{SF}^{(1)} -   \mathbf{V}_{SF}^{(1)} \boldsymbol{\Lambda}_F^{(0)} &= -\mathbf{C}_1 \mathbf{V}_{FF}^{(0)} 
\end{align}
\end{subequations}
From the first two subequations, we observe that the diagonal blocks of the first order perturbation are parallel to the unperturbed eigenvectors themselves. This perturbation can then be eliminated (or ignored) via a renormalization of the perturbed eigenvalue matrix. 
In contrast, the non-diagonal blocks require solving a Sylvester equation. In order to simplify the next order perturbation, we choose to work in the gauge where $\mathbf{V}_{SS}^{(1)} = \mathbf{V}_{FF}^{(1)} = 0$.
For the second order perturbation of $\boldsymbol{\Lambda}^{(2)}$, we find:
\begin{align}
    \mathbf{M}^{(0)} \mathbf{V}^{(2)} + \mathbf{M}^{(1)} \mathbf{V}^{(1)} &= \mathbf{V}^{(0)}\boldsymbol{\Lambda}^{(2)} + \mathbf{V}^{(1)}\boldsymbol{\Lambda}^{(1)} \nonumber \\&\quad +\mathbf{V}^{(2)}\boldsymbol{\Lambda}^{(0)}
\end{align}
Multiplying with $(\mathbf{V}^{(0)})^T$ from the left and substituting
$\boldsymbol{\Lambda}^{(1)} = 0$, we obtain
\begin{align}
    \boldsymbol\Lambda^{(2)} &= \text{diag} \left((\mathbf{V}^{(0)})^T \mathbf{M}^{(1)}\mathbf{V}^{(1)} \right) \nonumber \\
&= \text{diag} \left( \begin{bmatrix}
(\mathbf{V}_{SS}^{(0)})^T
    \mathbf{C}_1 \mathbf{V}_{FS}^{(1)} & \\ & (\mathbf{V}_{FF}^{(0)})^T
    \mathbf{C}_2 \mathbf{V}_{SF}^{(1)} \end{bmatrix} \right)
\end{align}
Finally, the second order eigenvector perturbation results in
\begin{subequations}
\label{eq:off_diag_perturb_second}
\begin{align}
\mathbf{S} \mathbf{V}_{SS}^{(2)} -   \mathbf{V}_{SS}^{(2)} \boldsymbol{\Lambda}_S^{(0)} &= \mathbf{V}_{SS}^{(0)} \boldsymbol{\Lambda}_S^{(2)} - \mathbf{C}_1 \mathbf{V}_{FS}^{(1)} \\
\mathbf{F} \mathbf{V}_{FF}^{(2)} -   \mathbf{V}_{FF}^{(2)} \boldsymbol{\Lambda}_F^{(0)} &=   \mathbf{V}_{FF}^{(0)} \boldsymbol{\Lambda}_F^{(2)} - \mathbf{C}_2 \mathbf{V}_{SF}^{(1)} \\
\mathbf{F} \mathbf{V}_{FS}^{(2)} -   \mathbf{V}_{FS}^{(2)} \boldsymbol{\Lambda}_S^{(0)} &= \mathbf{0} \\
\mathbf{S} \mathbf{V}_{SF}^{(2)} -   \mathbf{V}_{SF}^{(2)} \boldsymbol{\Lambda}_F^{(0)} &= \mathbf{0}
\end{align}
\end{subequations}
Here, since the spectra of $\mathbf{S}$ and $\mathbf{F}$ are guaranteed to be different (the spectrum of $\mathbf{F}$ is complex when $\kappa_f>0$ whereas the spectrum of $\mathbf{S}$ is real), the unique solution of the last two subequations is $\mathbf{V}_{SF}^{(2)} = \mathbf{V}_{FS}^{(2)} = \mathbf{0}$.
The magnitude of the second order perturbations in the system eigenvalues are thus given by the first row of $\mathbf{V}_{FS}^{(1)}$, up to a shift in the sign depending on the type of interaction with the filter.
From Eq. (\ref{eq:off_diag_perturb}), we find $\mathbf{V}_{FS}^{(1)}$ as (substituting $\mathbf{V}_{SS}^{(0)} = \mathbf{I}$):
\begin{align}
(\mathbf{V}_{FS}^{(1)})_{ki} &= \sum_j (\mathbf{V}_{FF}^{(0)})_{kj} \frac{((\mathbf{V}_{FF}^{(0)})^T \mathbf{C}_2 )_{ji}}{s_i - f_j}
\end{align}
where $s_i, f_i$ are the eigenvalues of $\mathbf{S}$, $\mathbf{F}$, respectively. $\mathbf{V}_{SF}^{(1)}$ can also be written similarly, 
\begin{align}
(\mathbf{V}_{SF}^{(1)})_{ki} &= \frac{( \mathbf{C}_1 \mathbf{V}_{FF}^{(0)} )_{ki}}{f_i - s_k}
\end{align}
Then, for the eigenvector perturbation to hold, we need to have
\begin{align}
    |s_i - f_j| \gg \tilde g \;\; \forall \;\;i, j.
\end{align}
Because the filter is designed for a flat response of bandwidth $B$, its $M$ modes have comparable linewidths, with no anomalously narrow mode. Hence, $\text{min}_j|\text{Im}[f_j]| \sim O(1/M^2)$, and  
$|s_i - f_j| \geq |\text{Im}[f_j]| \gtrsim 1/M^2$, and we need to have $\tilde g \ll 1/M^2$. Furthermore, for the second order perturbation analysis in the eigenvalues to hold, the perturbations need to be smaller than the spacing between the unperturbed eigenvalues, i.e.
\begin{align}
    |\Delta_i - \Delta_j| \gg  g^2/B, \;\; |f_i - f_j| \gg  g^2/B, \forall \, i,j.
\end{align}
If this condition is satisfied, the perturbative approach holds. The nonzero $\mathbf{V}_{SS}^{(2)}$ signifies that the different Gaussian modes hybridize when connected to the filter, which gives rise to a dressed eigenbasis. However, this is a second order effect. Each mode in the dressed eigenbasis experiences an effective dissipation, where the dissipation rate is $O(g^2/B)$, as seen from the second order correction in the eigenvalues.
In contrast, if there is a subspace of system modes with identical detunings, one needs to apply degenerate perturbation theory, where the correct eigenvector matrix corresponds to linear superpositions of the corresponding degenerate system modes.

\subsection{Without the Gaussian approximation}
\label{app_sub:non_gaussian}

Here, let us forego the Gaussian approximation on the adjacent transitions in the system mode, and consider the anharmonic oscillator with the Hamiltonian $\hat{H}_0 = -K \hat{d}^\dagger \hat{d}^\dagger \hat{d} \hat{d} + \omega_d \hat{d}^\dagger \hat{d}$. In order to approximate the effective dynamics, we will adiabatically eliminate the filter modes. For this purpose, we follow the formalism of Refs. \cite{reiter_effective_2012, chamberland_building_2022}. First, we define a \textit{ground subspace} with Hamiltonian $\hat{H}_g$, and an \textit{excited subspace} with Hamiltonian $\hat{H}_e$. The ground subspace describes the stable, non-decaying modes, whereas the excited subspace describes the excitations decaying back to the ground subspace.

For our filter system, given the small coupling rate $g \ll J_{i,i+1}$, $g \ll \kappa_f$, excitations in the filter modes will decay rapidly to the bath. Then, we can identify the ground subspace as the subspace where the filter is in its vacuum state $\ket{0_f}$. Similarly, the excited subspace can be identified as the subspace where the filter has an excitation. In both subspaces, the system mode can be in any state. To simplify the dynamics, let us go into an interaction frame with respect to the $\hat{H}_0$ of the system mode, and the bare Hamiltonian of the filter modes, $\sum_{i=1}^{M-1}\omega_f \hat{f}_i^\dagger \hat{f}_i$.
In this frame, we rewrite the Hamiltonian with the projection operators $\hat{P}_g\coloneqq\ket{0_f}\bra{0_f}$ and $\hat{P}_e\coloneqq I_f-\ket{0_f}\bra{0_f} \approx \sum_{k=1}^M \ket{e_k}\bra{e_k}$ with $\ket{e_k} \coloneqq \hat{f}_k^\dagger \ket{0_f}$ as
\begin{align}
    \hat{H} &= \begin{bmatrix}
        \hat{P}_g & \hat{P}_e
    \end{bmatrix} \begin{bmatrix}
         \hat{H}_g & \hat{V}_+ \\
         \hat{V}_- &  \hat{H}_e
    \end{bmatrix} \begin{bmatrix}
        \hat{P}_g \\ \hat{P}_e
    \end{bmatrix} \nonumber \\
    &= g \hat{f}_1^\dagger \left( \hat{d}(t)  \, e^{i (\omega_f+\omega_{\text{bs}}) t}  + \hat{d}^\dagger(t) \, e^{i (\omega_f+\omega_{\text{tm}}) t} \right)  \nonumber \\
    &\quad + \sum_{i=1}^{M-1} J_{i,i+1} \, \hat{f}_i \hat{f}_{i+1}^\dagger + \text{h.c.},
\end{align}
where $I_f$ is the identity operator on the filter modes, and $\hat{d}(t) = \sum_{k\geq 1} \sqrt{k}\ket{k-1}\bra{k}e^{-i(\omega_d - 2K(k-1))t}$ is the system mode in the interaction frame. Then, we can define the level-dependent detunings as in Eq. (\ref{eq:detunings_defn}). Doing so, we find
\begin{subequations}
\begin{align}
    \hat{H}_g &=  0 \\
    \hat{H}_e &=  I_s \otimes\left(\sum_{i=1}^{M-1} J_{i,i+1} \, \ket{e_i}\bra{e_{i+1}} + \text{h.c.}\right)  \\
    \hat{V}_+ &= g \sum_{k\geq 1} \sqrt{k}\left(\ket{k-1}\bra{k}e^{i \Delta_{\text{bs},k} t} + \ket{k}\bra{k-1}  e^{i \Delta_{\text{tm},k} t} \right) \nonumber \\ &\quad \otimes \ket{e_1}\bra{0_f}
\end{align}
\end{subequations}
$\hat{V}_- = \hat{V}_+^\dagger$, and
$\ket{e_1} = \hat{f}_1^\dagger \ket{0_f}$ is a single excitation on the first filter mode. The operator $\hat{V}_+$ describes the transition from the ground subspace to the excited subspace. Let us denote the separate time-dependent terms as
\begin{align}
    \hat{V}_+ = \sum_{x \in T}\sum_{k \geq 1} \hat{V}^x_{+,k} e^{i\Delta_{x, k} t} 
\end{align}
with $T = \{\text{bs}, \text{tm}\}$, $\hat{V}^\text{bs}_{+,k} = g  \sqrt{k}\ket{k-1}\bra{k}\otimes \ket{e_1}\bra{0_f}$, $\hat{V}^\text{tm}_{+,k} = g  \sqrt{k}\ket{k}\bra{k-1}\otimes \ket{e_1}\bra{0_f}$, and $\hat{V}^x_{-,k} = (\hat{V}^x_{+,k})^\dagger$.
We similarly write the Lindblad operator $ \mathcal{D}[\sqrt{\kappa_f}\hat{f}_M]$ as
$\hat{L} = \sqrt{\kappa_f} \ket{0_f}\bra{e_M}$. Then, the unitary evolution in the excited subspace described by $\hat{H}_e$ is combined with the Lindblad operator to form a non-Hermitian operator that describes the full dynamics in this subspace,
\begin{align}
    \hat{H}_\text{NH} &= \hat{H}_e - \frac{i}{2} \hat{L}^\dagger \hat{L} \nonumber \\
    &= I_s \otimes\left(\sum_{i=1}^{M-1} J_{i,i+1} \, \ket{e_i}\bra{e_{i+1}} + \text{h.c.} \right) \nonumber \\&\quad - I_s \otimes\frac{i \kappa_f}{2} \ket{e_M}\bra{e_M}
\end{align}
Given this operator, the effective Hamiltonian in the ground subspace is found as
\begin{align}
    \hat{H}_\text{eff} &= -\frac{1}{2} \sum_{x,y \in T}\sum_{k,n \geq 1} \hat{V}_{-,k}^x \left[ \hat{F}_{x, k} + \hat{F}_{y, n} ^\dagger\right] \nonumber \\
    &\quad\quad\hat{V}_{+,n}^{y} e^{i(\Delta_{y,n} - \Delta_{x,k})t}
\end{align}
with $\hat{F}_{x, k} \coloneqq (\hat{H}_\text{NH} - \Delta_{x,k})^{-1} $.
The sum over the bs-bs and tm-tm terms in $\hat{V}_+$ and $\hat{V}_-$ will give rise to time-independent contributions, whereas the bs-tm cross terms will oscillate with frequencies $|\Delta_{\text{bs}, k} - \Delta_{\text{tm}, k-1}|, \; k\geq 2$. The time-dependent terms can be dropped with the RWA since $|\Delta_{\text{bs}, i}-\Delta_{\text{tm}, k}| \sim 2\omega_d \gg g^2/\kappa_f$. This approximation will result in
\begin{align}
    \hat{H}_\text{eff} &\approx -\frac{g^2}{2} \sum_{k\geq1} k \left[ \bra{e_1} \hat{F}_{\text{bs}, k} + \hat{F}_{\text{bs}, k}^\dagger \ket{e_1}   \ket{k}\bra{k} + \right. \nonumber \\
    &\left. \quad \bra{e_1} \hat{F}_{\text{tm}, k} + \hat{F}_{\text{tm}, k}^\dagger \ket{e_1}   \ket{k-1}\bra{k-1} \right] \otimes \ket{0_f} \bra{0_f}
\end{align}
Thus, the interaction with the excited subspace results in an $O(g^2)$ perturbation in the energy levels of the system in the ground subspace, and leaves the eigenbasis unchanged as cross terms within the same channel (bs-bs or tm-tm with $k \neq l$) vanish identically by the orthogonality of the projectors $\ket{k}\bra{k-1}$ for different $k$. Similarly, we can write the effective Lindblad operator in this space as
\begin{subequations}
\begin{align}
    &\hat{L}_{\text{eff}}(t) =
   \hat{L} \sum_{x \in T}\sum_{k \geq 1} \hat{F}_{x,k} \hat{V}^x_{+,k} e^{i\Delta_{x, k} t} \nonumber \\
    &\quad\quad\quad=\sum_{k\geq 1} \left( r_k^\text{bs} \hat{L}_k\, e^{i\Delta_{\text{bs},k} t} + r_k^\text{tm}\hat{L}_k ^\dagger  e^{i\Delta_{\text{tm},k} t}  \right), \\
    & \hat{L}_k := g\sqrt{\kappa_f}\,\sqrt{k}\, \ket{k-1}\bra{k},
\end{align}
\end{subequations}
where 
$r_k^x := \bra{e_M}\hat{F}_{x,k}\ket{e_1}$ .
Then, we observe that in the ground subspace, interacting with the filter with the beam splitter (two-mode squeezing) interaction will cause an effective Lindblad operator $\propto \hat{d}$ ($\propto \hat{d}^\dagger$), resulting in engineered decay (gain). 
The engineered dissipation rates are proportional to $g^2$, the filter interaction rate to the second power. Finally, let us show the approximate independence of the engineered dissipation on each transition $\ket{k} \leftrightarrow\ket{k+1}$. The effective Lindblad master equation in the ground subspace can be written as
\begin{align}
\label{app_eq:master_eqn_before_rwa}
    \dot{\hat{\rho}} = -i[\hat{H}_\text{eff}, \hat{\rho}] + \hat{L}_{\text{eff}} \hat{\rho} \hat{L}_{\text{eff}}^\dagger -\frac{1}{2} \{ \hat{L}_{\text{eff}}^\dagger \hat{L}_{\text{eff}}, \hat{\rho} \}
\end{align}
Here, the terms oscillating at different frequencies in $\hat{L}_{\text{eff}}(t)$ will give rise again to oscillating terms. However, now, there will be bs-bs and tm-tm cross terms as well.
The RWA can be applied if, for all $k,n$,
\begin{align}
\label{app_eq:rwa_cond}
    |\Delta_{\text{x}, k}-\Delta_{\text{x}, n}| \gg g^2\kappa_f\sqrt{kn}|r_{k}^xr_{n}^x|,
\end{align}
for $x \in T$. From the tridiagonal structure of $\hat{H}_\text{NH}$, $|r_{n}^x|$ is found exactly as
\begin{align}
    |r_{n}^x| = \frac{\prod_{i=1}^M J_i}{|\text{det}(\hat{H}_\text{NH}-\Delta_{x,n})|}
\end{align}
assuming that $J_i >0 \; \forall \,i$. For an upper bound on the required separation of the detunings, we want to estimate an upper bound on the r.h.s. of this inequality, i.e. an upper bound on the obtainable effective decay rate obtained with the filter. For this purpose, assuming that $\{J_i\}, \kappa_f$ is such that the filter response is almost flat, we can set $\Delta_{x,k}=0$ (i.e. the transition is in the middle of the filter). Then, $|r_{n}^x|$ is given by
\begin{align}
    |r_{n}^x|= \begin{cases}
    \frac{\prod_{i=1}^M J_i}{\prod_{i=0}^{(M-2)/2} J_{2i+1}^2} & \text{if} \;$M$ \;\text{is even} \\[3mm]
  \frac{2\prod_{i=1}^M J_i}{\kappa_f\prod_{i=0}^{(M-3)/2} J_{2i+1}^2} & \text{if} \;$M$ \;\text{is odd}
    \end{cases}
\end{align}
For a bandpass filter, we will have that $J_i \sim \kappa_f \sim O(B)$: thus, $|r_{n}^x| \sim O(1/B)$, and the condition in Eq. (\ref{app_eq:rwa_cond}) reduces to
\begin{align}
    |\Delta_{\text{x}, k}-\Delta_{\text{x}, n}| \gg \frac{g^2 \sqrt{kn}}{B}
\end{align}
If this condition is satisfied for $x \in T$, we can rewrite Eq. (\ref{app_eq:master_eqn_before_rwa}) as
\begin{align}
\label{eq:app_effective_master_eqn}
     \dot{\hat{\rho}} = -i[\hat{H}_\text{eff}, \hat{\rho}] + \sum_{k\geq 1} \left( |r_k^\text{bs}|^2  \mathcal{D}[\hat{L}_k] \hat{\rho} + |r_k^\text{tm}|^2  \mathcal{D}[\hat{L}_k^\dagger] \hat{\rho} \right)
\end{align}
The coherent off-diagonal terms dropped by the RWA in this equation correspond to the non-zero $\mathbf{V}_{SS}^{(2)}$ perturbation retained in the Gaussian approximation of Appendix \ref{app_sub:gaussian}. Both are controlled by the same small parameter $g^2/(B|\Delta_i-\Delta_j|)$, and do not affect the leading-order engineered dissipation rates. 
Thus, the effective dynamics can be approximated with independent jump operators $g^2\kappa_f k|r_k^\text{bs}|^2\mathcal{D}[\ket{k-1}\bra{k}]$ describing engineered decay, and $g^2\kappa_f k|r_k^\text{tm}|^2\mathcal{D}[\ket{k}\bra{k-1}]$ describing engineered gain, $k \geq 1$. Lastly, the engineered dissipation rates can be suppressed with a far-detuned interaction (i.e. large $\Delta_{x,k}$), since $|r_k^x| \ll 1$ if $\Delta_{x,k}\gg J_{i,i+1}$, $\Delta_{x,k}\gg \kappa_f$.


\section{Gates between 0-n Fock qubits}
\label{app:fock_qubit_gates}

In this Appendix, we consider the task of implementing bias-preserving entangling gates between 0-n Fock qubits. We sketch proof-of-principle implementations of bias-preserving CZ and CX gates between Fock qubits, demonstrating that such operations can in principle be realized with the same dispersive interactions used in the main text. While the CZ implementation is straightforward, the proposed CX gate is considerably more involved than the Fock-cat CX gate in the main text. We emphasize these proof-of-principle implementations are not practically optimized, the relative complexity of implementing a bias-preserving CX gate between Fock qubits underscores the practical appeal of the hybrid cat-Fock architecture described in the main text.

\subsection{CZ gate}

In the Fock basis, a CZ gate between 0-n Fock qubits corresponds to the unitary
\begin{equation}
    \mathrm{CZ} = \ket{0,0}\bra{0,0}+\ket{0,n}\bra{0,n}+\ket{n,0}\bra{n,0} - \ket{n,n}\bra{n,n}.
\end{equation}
This gate can be straightforwardly implemented with a dispersive coupling between 0-n Fock qubits. The dispersive interaction $H = \chi \hat{a}_1^\dagger \hat{a}_1 \hat{a}_2^\dagger \hat{a}_2$ between two modes $\hat{a}_1$ and $\hat{a}_2$ directly corresponds to a $ZZ$ interaction between 0-n Fock qubits supported in these modes. Evolution under this interaction for a time $t = \pi/(n^2 \chi)$ implements the CZ gate. The dispersive coupling strength $\chi$ can be dynamically controlled using a tunable coupler~\cite{yan2018tunable,stehlik2021tunable}, enabling the gate to be turned on and off at will. The implementation is bias-preserving for the same reason as the cat-Fock CX gate in the main text: both consist of short dispersive evolutions after which up to $\lceil n/2 \rceil - 1$ intrinsic decay or heating events can be corrected by the engineered dissipation. 
\subsection{CX gate}

In the Fock basis, a CX gate between 0-n Fock qubits corresponds to the unitary
\begin{equation}
    \mathrm{CX} = \ket{0,0}\bra{0,0}+\ket{0,n}\bra{0,n}+\ket{n,n}\bra{n,0} + \ket{n,0}\bra{n,n}.
\end{equation}
This gate is less straightforward to implement than the CZ gate, owing to the need to implement controlled multiphoton transitions $\ket{0} \leftrightarrow \ket{n}$ in a bias-preserving way. Nevertheless, we show below that this operation can be implemented through a combination of dispersive couplings, single-photon drives, and engineered dissipation. We emphasize that this construction is a proof of principle and is not practically optimized.

\begin{figure*}
    \centering
\includegraphics[width=0.95\linewidth]{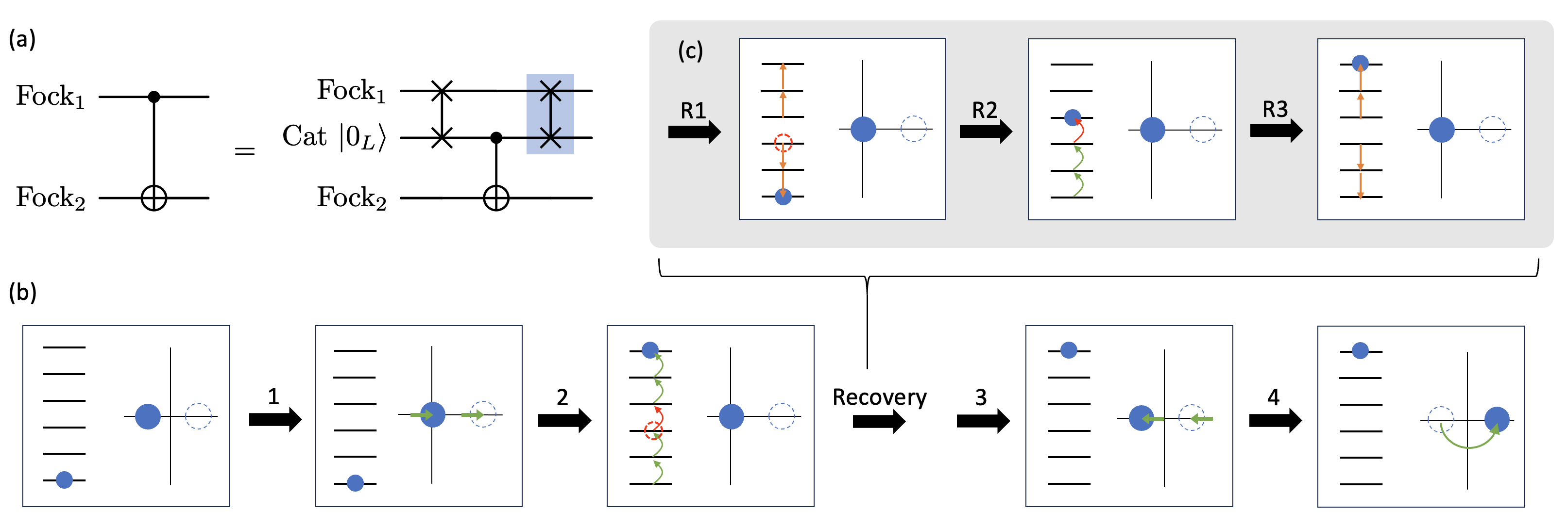}
    \caption{Proof-of-principle implementation of a bias-preserving CX between 0-n Fock qubits. (a) Circuit decomposition. The CX can be implemented by transferring the state of one Fock qubit into an auxiliary cat qubit, then using Fock-cat CX from the main text. Here state transfers are represented by SWAP gates, and the cat is initialized in $|0_L\rangle \approx |\alpha\rangle_C$. (b) State transfer sequence. The pulse sequence implementing the highlighted SWAP gate is depicted step by step, illustrating each step's action on the initial state $|{-}\alpha\rangle_C |0\rangle_F$.  Green arrows indicate coherent unitary operations, and orange arrows indicate engineered dissipation. The red arrow and dashed red circle indicate a possible pulse error during step 2 and resulting state of the Fock qubit. (c) Recovery sequence. Pulse errors like that indicated in red in (b) are corrected by the recovery sequence, where steps R2 and R3 can be repeated to ensure a sufficiently high noise bias is maintained. }
\label{fig:Fock_CX}
\end{figure*}

To simplify the implementation, we consider a CX gate mediated by an auxiliary cat qubit. This allows us to leverage the Fock-controlled CX gate from the main text as a subroutine. The protocol proceeds as follows (see Fig.~\ref{fig:Fock_CX}): the state of one Fock qubit is first transferred to the cat qubit, a Fock-controlled CX gate generates entanglement with the second Fock qubit, and finally the cat's state is transferred back to the first Fock qubit. Thus, to implement a bias-preserving CX gate between Fock qubits, we only additionally require bias-preserving state transfer between a Fock qubit and a cat qubit. 

\subsubsection{State transfer}

Inspired by the cat-transmon gate constructions in Ref.~\cite{hann_hybrid_2025}, these state transfer operations are implemented using composite pulse sequences involving displacements of the cat mode and photon-number-selective driving. We first describe a simplified version of the sequence to illustrate the key mechanism, then show how it can be augmented to preserve noise bias.

We focus on the transfer from the cat qubit to the Fock qubit; the reverse transfer can be implemented analogously by reversing the sequence. We denote the cat and Fock qubit states with subscripts $_C$ and $_F$, respectively. Starting from the initial state $(c_0\ket{-\alpha}_C + c_1\ket{\alpha}_C)\ket{0}_F$ for the joint cat-Fock system:
\begin{enumerate}
    \item The cat mode is displaced by $+\alpha$: \\$\to (c_0\ket{0}_C + c_1\ket{2\alpha}_C)\ket{0}_F$.
    \item A cascaded sequence of $n$ vacuum-selective $\pi$-pulses~\cite{krastanov2015universal, heeres2015cavity} is applied to the Fock qubit, sequentially driving $\ket{0}_F\to\ket{1}_F\to\cdots\to\ket{n}_F$, but only when the cat mode is in vacuum: \\$\to c_0\ket{0}_C\ket{n}_F + c_1\ket{2\alpha}_C\ket{0}_F$.
    \item The cat mode is displaced by $-\alpha$: \\$\to c_0\ket{-\alpha}_C\ket{n}_F + c_1\ket{\alpha}_C\ket{0}_F$.
    \item A Fock-controlled CX is applied: \\$\to \ket{\alpha}_C(c_0\ket{n}_F + c_1\ket{0}_F)$.
\end{enumerate}
With this sequence, information initially encoded in the cat qubit is transferred to the Fock qubit as desired. The key steps are step~2, where the Fock qubit is entangled with the cat using vacuum-selective pulses, and step~4, where the Fock-controlled CX gate from the main text disentangles the cat qubit.

\subsubsection{Preserving noise bias via a recovery sequence}

Among the primitive operations involved, the Fock-controlled CX gate (step 4) and displacements (steps 1, 3) are natively bias-preserving. Recall the bias-preserving nature of the Fock-controlled CX is established in the main text. The only operations requiring additional care are the vacuum-selective pulses on the Fock qubit, which we address in detail below.

The simplified pulse sequence above is not bias-preserving due to the possibility of pulse errors in step~2. An error in the selective pulses can leave the joint system in the state $c_0\ket{0}_C\ket{m}_F + c_1\ket{2\alpha}_C\ket{0}_F$ for some $m \in [0,n)$, depending on which pulse suffers the error. For $m < (n+1)/2$, this would correspond to a bit flip on the Fock qubit after stabilization.

To correct such errors, we insert a recovery sequence between steps~2 and~3:
\begin{enumerate}
    \item[R1.] Engineered dissipation is applied to the Fock qubit, mapping the erroneous state to:
    \begin{equation}
        \begin{cases}
            c_0\ket{0}_C\ket{0}_F + c_1\ket{2\alpha}_C\ket{0}_F & \text{if } m < (n+1)/2, \\
            c_0\ket{0}_C\ket{n}_F + c_1\ket{2\alpha}_C\ket{0}_F & \text{if } m \geq (n+1)/2.
        \end{cases}
    \end{equation}
    \item[R2.] A sequence of $(n+1)/2$ vacuum-selective $\pi$-pulses is applied to the Fock qubit, driving $\ket{0}_F\to\ket{1}_F\to\cdots\to\ket{(n+1)/2}_F$:
    \begin{equation}
        \begin{cases}
            c_0\ket{0}_C\ket{(n+1)/2}_F + c_1\ket{2\alpha}_C\ket{0}_F & \text{if } m < (n+1)/2, \\
            c_0\ket{0}_C\ket{n}_F + c_1\ket{2\alpha}_C\ket{0}_F & \text{if } m \geq (n+1)/2.
        \end{cases}
    \end{equation}
    \item[R3.] Engineered dissipation is applied again:
    \begin{equation}
        c_0\ket{0}_C\ket{n}_F + c_1\ket{2\alpha}_C\ket{0}_F \quad \text{in both cases.}
    \end{equation}
\end{enumerate}

This sequence effectively realizes a cat-state-dependent dissipation on the Fock qubit. When the cat mode is in $\ket{2\alpha}_C$, the vacuum-selective pulses do not act, and the sequence reduces to the usual Fock-qubit dissipation. When the cat mode is in vacuum, the sequence maps all Fock states $\ket{m}_F \to \ket{n}_F$ for any $m \in [0,n]$, correcting any pulse error from step~2.

The recovery sequence itself involves vacuum-selective pulses (step R2) that can also suffer errors. Such errors are corrected by repeating steps R2 and R3. Since a pulse error must occur during \emph{each} repetition to induce a bit-flip error, $k < n/2$ repetitions suppress the bit-flip probability due to pulse errors to $\mathcal{O}(p_e^{k+1})$, where $p_e$ is the single-pulse error probability. With sufficiently many repetitions, pulse errors can therefore be exponentially suppressed, preserving the Fock qubit's noise bias.

\section{CX bit-flip probability}
\label{app:bit_flip_calcs}

Here, the logical states of the data qubits are defined from $\ket{\pm} = \mathcal{N}_{\pm} (\ket{\alpha} \pm \ket{-\alpha})$, where $\mathcal{N}_{\pm} = (2 \pm 2e^{-2|\alpha|^2})^{-1/2}$. These states are perfectly orthogonal to each other, $\braket{+}{-} = 0$. The logical states are 
\begin{align}
    \ket{0_L} &= \frac{\ket{+} + \ket{-}}{\sqrt{2}}, \quad
    \ket{1_L} = \frac{\ket{+} - \ket{-}}{\sqrt{2}} 
\end{align}
Writing the logical states in the coherent state basis, we obtain
\begin{align}
    \ket{k_L} = \ket{\alpha} \frac{(\mathcal{N}_+ + (-1)^k \mathcal{N}_-)}{\sqrt{
    2}} + \ket{-\alpha}\frac{(\mathcal{N}_+ -(-1)^k \mathcal{N}_-)}{\sqrt{
    2}}
\end{align}
with $k =0,1$. For $\alpha \gg 1$, $\ket{0_L} \approx \ket{\alpha}$, and $\ket{1_L} \approx \ket{-\alpha}$. 

Let us assume a control qubit is prepared in $\ket{\psi} = \ket{m}$ in the Fock basis. We apply the dispersive coupling Hamiltonian $H = \chi_0 (\hat{a}^\dagger \hat{a}) \otimes (\hat{d}^\dagger \hat{d})$, where $\hat{a}$ ($\hat{d}$) is the annihilation operator of the control (data) qubit. Initializing the data qubit in $\ket{0_L}$, we apply this Hamiltonian on the state $\ket{m} \otimes \ket{0_L}$ for a time $t$. Given that we have
\begin{align}
    e^{-i\chi_0 t (\hat{a}^\dagger \hat{a}) \otimes (\hat{d}^\dagger \hat{d})} \ket{m} \otimes \ket{\alpha}
    &=\ket{m} \otimes \ket{\alpha e^{-i \chi_0 m t}},
\end{align}
this interaction will result in the following state:
\begin{align}
\label{eq:control_data_qubit_after_rotation}
    \ket{m} \otimes \left( \ket{\tilde\alpha} \frac{(\mathcal{N}_+ + \mathcal{N}_-)}{\sqrt{
    2}} + \ket{-\tilde\alpha}\frac{(\mathcal{N}_+ -\mathcal{N}_-)}{\sqrt{
    2}}  \right)
\end{align}
Here, $\tilde\alpha = \alpha e^{-i \theta}$, and $\theta \coloneqq  \chi_0 mt$ is the rotation angle. We then trace the control qubit out and dissipatively stabilize the data qubit such that it is contained in the logical manifold $\{\ket{0_L}, \ket{1_L}\}$. The stabilization is performed with a two-photon process. The initial state can be written as
\begin{align}
    \hat{\rho}_i(\theta) &= a_+^2 \ket{\tilde\alpha}\bra{\tilde\alpha} + a_-^2 \ket{-\tilde\alpha}\bra{-\tilde\alpha} \nonumber \\
    & \quad + a_+a_-(\ket{\tilde\alpha}\bra{-\tilde\alpha}+\ket{-\tilde\alpha}\bra{\tilde\alpha})
\end{align}
where $a_\pm = \frac{(\mathcal{N}_+ \pm \mathcal{N}_-)}{\sqrt{2}}$. The evolution of the dissipatively stabilized data qubit will obey
\begin{align}
    \frac{d\hat{\rho}(t)}{dt} = \kappa_2 \mathcal{D}[\hat{d}^2 - \alpha^2]\hat{\rho}(t)
\end{align}
where $\kappa_2$ is the engineered dissipation rate, and $\mathcal{D}[\hat{A}] \coloneqq \hat{A}\hat{\rho} \hat{A}^\dagger - \frac{1}{2} \{\hat{A}^\dagger \hat{A}, \hat{\rho}\}$. The asymptotic steady state at the end of the stabilization is in the form of
\begin{align}
\label{eq:rho_infinity}
    \hat{\rho}_\infty &= c_{++} \ketbra{+}{+} + (1-c_{++}) \ketbra{-}{-} \nonumber \\
    &\quad + c_{+-} \ketbra{+}{-} + c_{+-}^* \ketbra{-}{+}.
\end{align}
We can write \cite{mirrahimi_dynamically_2014}
\begin{align}
\label{eq:c++c+-_exact}
    c_{++} &= 2 \mathcal{N}_+^2\frac{1+e^{-2|\alpha|^2}}{2} = \frac{1}{2},\\ c_{+-} &= \frac{2i \mathcal{N}_+ \mathcal{N}_- |\alpha|^2 }{\sqrt{e^{4|\alpha|^2}-1}} F_\alpha(\theta) = \frac{i  |\alpha|^2 e^{-2|\alpha|^2}}{1-e^{-4|\alpha|^2}} F_\alpha(\theta)
\end{align}
where we have defined
\begin{align}
    F_\alpha(\theta) \coloneqq \int_0^\pi d\phi \,e^{-i(\phi-\theta)} I_0(2|\alpha|^2 \sin{(\phi- \theta)})
\end{align}
Here, $I_0$ is the modified Bessel function of the first kind. $c_{++}, c_{+-}$ characterize the final state exactly. Let us now derive the bit-flip probability if a CX gate is applied to two data qubits, conditioned on the Fock state $\ket{m}$. We refer to this probability as $\tilde p_\text{bf}(\theta)$. First, defining the joint state of the control and the data qubit after a rotation, given in Eq. (\ref{eq:control_data_qubit_after_rotation}), as $\ket{m} \otimes \ket{\xi}$, we will now have a state $\ket{m} \otimes \ket{\xi} \otimes \ket{\xi}$. Tracing out the control qubit, the data qubits will be in the form of $\hat{\rho}_\infty \otimes \hat{\rho}_\infty$, where $\hat{\rho}_\infty$ is characterized in Eqs. (\ref{eq:rho_infinity}), (\ref{eq:c++c+-_exact}). The bit-flip probability can then be written as
\begin{align}
    \tilde{p}_\text{bf}(\theta) &= 1 - (\bra{0_L} \hat{\rho}_\infty \ket{0_L})^2 - (\bra{1_L} \hat{\rho}_\infty \ket{1_L})^2 \nonumber \\
    &=\frac{1}{2} - 2 \text{Re}[c_{+-}]^2 \nonumber \\
    &= \frac{1}{2} \left[ 1 - \left( \frac{  |\alpha|^2}{\sinh{(2|\alpha|^2)}}  \text{Re}[F_\alpha(\theta)]  \right)^2 \right]
\end{align}
From the definition of the Bessel function, $\frac{  |\alpha|^2}{\sinh{(2|\alpha|^2)}}  \text{Re}[F_\alpha(\theta)]$ can be rewritten as the following for $0 \leq \theta \leq \pi$:
\begin{align}
\label{eq:I_theta_calculations}
    &\frac{|\alpha|^2}{\sinh{(2|\alpha|^2)}} \sum_{m=0}^\infty \frac{|\alpha|^{4m}}{(m!)^2} \int_0^\pi d\phi \,\sin^{2m+1}{\left(\phi-\theta \right)} \nonumber \\
    &= 1  -  \frac{2|\alpha|^2}{\sinh{(2|\alpha|^2)}} \sum_{m=0}^\infty \frac{|\alpha|^{4m}}{(m!)^2} \int_0^{\theta} d\phi \, \sin^{2m+1}{(\phi)} \nonumber \\
    &=  1 -  \frac{2|\alpha|^2}{\sinh{(2|\alpha|^2)}} \int_0^{\theta} d\phi \, \sin{(\phi)} I_0\left(2|\alpha|^2\sin{(\phi)} \right) \nonumber \\
    &\coloneqq 1 - M_\alpha(\theta)
\end{align}
which gives the expression in Eq. (\ref{eq:bit_flip_probability_defn}) in the main text.

Given a control qubit in the state of Eq. (\ref{eq:control_qubit_imperfect_pulse_final_state}) in the main text, let us compute the total bit-flip probability from the CX gate on two data qubits. We again assume dispersive coupling between the control and the data qubit. Notice that the terms that are non-diagonal in the Fock basis in Eq. (\ref{eq:control_qubit_imperfect_pulse_final_state}) will lead to zero contribution in the final state of the data qubits, when the control qubit is traced out. Then, the initial state of the data qubits before the stabilization is written as
\begin{align}
    &\frac{1}{2} [(1+p_e) \hat{\rho}_i(0) \otimes \hat{\rho}_i(0)  + (1-p_e)^n \hat{\rho}_i(n \theta) \otimes \hat{\rho}_i(n \theta)   \nonumber \\
    & \quad + p_e \sum_{j=1}^{n-1} (1-p_e)^{j}  \hat{\rho}_i(j \theta) \otimes \hat{\rho}_i(j \theta) ]
\end{align}
where $\theta$ is the rotation angle, and $1\geq p_e \geq 0$ is the pulse error probability. We set $n \theta = \pi$, such that the data qubits undergo a $\pi$-pulse $\ket{0_L} \rightarrow \ket{1_L}$ if the control qubit is in state $\ket{n}$. 
The first two terms do not contribute to the bit-flip probability after the stabilization (since $p_\text{bf}(0) = p_\text{bf}(\pi) = 0$). Then, Eq. (\ref{eq:total_bit_flip_prob}) in the main text is found to be
\begin{align}
    p_\text{bf} = \frac{p_e}{2}  \sum_{k=1}^{n-1} (1-p_e)^k \tilde{p}_\text{bf}\left(\theta = \frac{\pi k}{n} \right)
\end{align}
Linearizing for $p_e \ll 1$, we find that $p_\text{bf} \approx \frac{p_e}{2} \sum_{k=0}^{n} \tilde{p}_\text{bf}\left(\theta = \frac{\pi k}{n} \right)$. Here, note that we extended the summation limits to include $k=0$ and $k=n$, as their contribution to the sum is zero. Using the definition for $p_\text{bf}(\theta)$ in Eq. (\ref{eq:bit_flip_probability_defn}) of the main text, we expand this expression as
\begin{align}
\label{eq:bit_flip_total_M}
     p_\text{bf} &\approx \frac{p_e}{2} \sum_{k=0}^n \left[M_\alpha \left(\frac{\pi k}{n}\right) - \frac{1}{2} M_\alpha^2\left(\frac{\pi k}{n}\right) \right]\nonumber \\
     &\approx \frac{p_e n}{2} \int_0^1 dz \left[M_\alpha \left(\pi z\right) - \frac{1}{2} M_\alpha^2\left(\pi z\right) \right] 
\end{align}
where we converted the summation over the Fock levels to an integral in the limit that $n \gg 1$.
From the symmetry of $M(\pi z)$, we have that $M_\alpha(\pi z) = 2 - M_\alpha(\pi(1-z))$ for $z \in [\frac{1}{2}, 1]$. Then, we find:
 \begin{align}
     &\int_0^1 dz \, M_\alpha \left(\pi z\right) \nonumber \\& =  \int_0^\frac{1}{2} dz \, M_\alpha \left(\pi z\right) + \int_\frac{1}{2}^1 dz \, \left[2-M_\alpha \left(\pi (1-z)\right) \right] \nonumber \\
     &= 1
 \end{align}
We can similarly rewrite the second term in Eq. (\ref{eq:bit_flip_total_M}) as
 \begin{align}
 \label{eq:M_alpha_sq}
     &\int_0^1 dz \, M_\alpha^2 \left(\pi z\right) \nonumber \\& =  \int_0^\frac{1}{2} dz \, M^2_\alpha \left(\pi z\right) + \int_\frac{1}{2}^1 dz \, \left[2-M_\alpha \left(\pi (1-z)\right) \right]^2 \nonumber \\
     &= 2-4 \int_0^\frac{1}{2} dz \, M_\alpha \left(\pi z\right) +2 \int_0^\frac{1}{2} dz \, M_\alpha^2 \left(\pi z\right)
 \end{align}
 In order to approximate this expression, we notice that the integrand in the definition of $M_\alpha(\theta)$ in Eq. (\ref{eq:I_theta_calculations}) is sharply peaked around $\pi/2$. In the vicinity of $\pi/2$, we can approximate $I_0(2|\alpha|^2\sin{\phi}) \approx \exp(2|\alpha|^2\sin{\phi})/\sqrt{4\pi |\alpha|^2\sin{\phi}}$, and $\sin{(\phi)}\approx 1-\frac{(\frac{\pi}{2}-\phi)^2}{2}$. Then, using the Laplace method, we rewrite $M_\alpha(\theta)$ as
 \begin{align}
     M_\alpha(\theta) \approx \frac{|\alpha|/\sqrt{\pi}}{\sinh{(2|\alpha|^2)}} \int_{-\pi/2}^{\theta-\pi/2} du \,  e^{2|\alpha|^2 (1-\frac{u^2}{2})} \nonumber \\
     \approx \frac{|\alpha|/\sqrt{\pi}}{\sinh{(2|\alpha|^2)}} \int_{-\infty}^{\theta-\pi/2} du \,  e^{2|\alpha|^2 (1-\frac{u^2}{2})}
 \end{align}
For $0\leq \theta \leq \pi/2$, we can further write
\begin{align}
    M_\alpha(\theta) \approx 1 - \text{erf}\left(|\alpha|\left( \frac{\pi}{2} -\theta\right)\right),
\end{align}
where $\text{erf}(z)$ is the erf function. Finally, plugging this into Eq. (\ref{eq:M_alpha_sq}), we obtain
\begin{align}
    \int_0^1 dz \, M_\alpha^2 \left(\pi z\right) &\approx 1 + 2 \int_0^\frac{1}{2} dz \, \text{erf}^2\left(|\alpha|\pi z\right) \nonumber \\
    &\approx 2 - \left(\frac{2}{\pi}\right)^{3/2}\frac{1}{|\alpha|}
\end{align}
for $|\alpha|\gg 1$. Thus, Eq. (\ref{eq:bit_flip_total_M}) can finally be approximated as $p_\text{bf} \approx \frac{p_en}{\sqrt{2 \pi^3}|\alpha|}$ for $|\alpha|, n \gg 1$.

\section{Bit-flip floor}
\label{app:bit_flip_floor}

We now derive the bit-flip floor analytically. We assume the ideal setting of the 0-n Fock qubit where we engineer dissipation with rate $\kappa_s$ between adjacent levels in the lower and upper half spaces. The transition rates are summarized in Eq. (\ref{eq:idealized_rates}), and we work in the regime where intrinsic loss processes are much weaker than the engineered dissipation $\kappa_i^{\uparrow, \downarrow} \ll \kappa_s$, such that $r_{\uparrow, \downarrow} \coloneqq \kappa_i^{\uparrow, \downarrow}/\kappa_s \ll 1$. 

In this setting, we observed that the system stabilizes to a metastable state in short time-scales $t \sim 1/\kappa_s$. As the populations in each level converge to their steady state values, the CX bit-flip probability when the qubit is used as the control qubit also converges to a fixed value. We refer to this probability as the \textit{bit-flip floor}. Let us analyze how the bit-flip floor varies as a function of the number of levels in the qubit $N$, and the mean photon number in the cat qubit $\alpha$. Furthermore, let us examine the speed of convergence to the bit-flip floor. In the following, we assume that the Fock qubit is prepared with pulses with error probability $p_e$, as described in Subsection \ref{subsec:bf_prob_before_stabilization}. Thus, the initial state at time $t=0$ is given in Eq. (\ref{eq:control_qubit_imperfect_pulse_final_state}).

Fixing $|\alpha|^2 = 10$, $p_e = 10^{-3}$, $r_\downarrow = 10^{-3}$, and $r_\uparrow = 10^{-5} $, we plot the total bit-flip probability as a function of the stabilization time $t$ for various $n$ in Fig \ref{fig:bit_flip_floor_a}. From the Figure, we observe that the bit-flip probability decays exponentially with time before converging to the bit-flip floor.
Then, we plot the scaling of the bit-flip floor with respect to the number of levels $N$ for various $|\alpha|$ in Fig \ref{fig:bit_flip_floor_b}. We fix 
$\kappa_i^\uparrow = 10^{-2}\kappa_i^\downarrow$ and $p_e = 10^{-3}$. We plot the case where $r_\downarrow = 10^{-3}$ with the circles, and $r_\downarrow = 10^{-4}$ with the stars.  
For relatively small $N$, the bit-flip floor decays exponentially with $N$, where we observe that the decay rate is proportional to $|\alpha|^2$. As $N \gg 1$, the scaling is modified into a power law. Let us derive this scaling analytically, assuming that the limiting loss process for the bit-flip floor is intrinsic decay, i.e. $\kappa_i^\downarrow \gg \kappa_i^\uparrow$.

\begin{figure}
    \centering
    \includegraphics[width=0.85\linewidth]{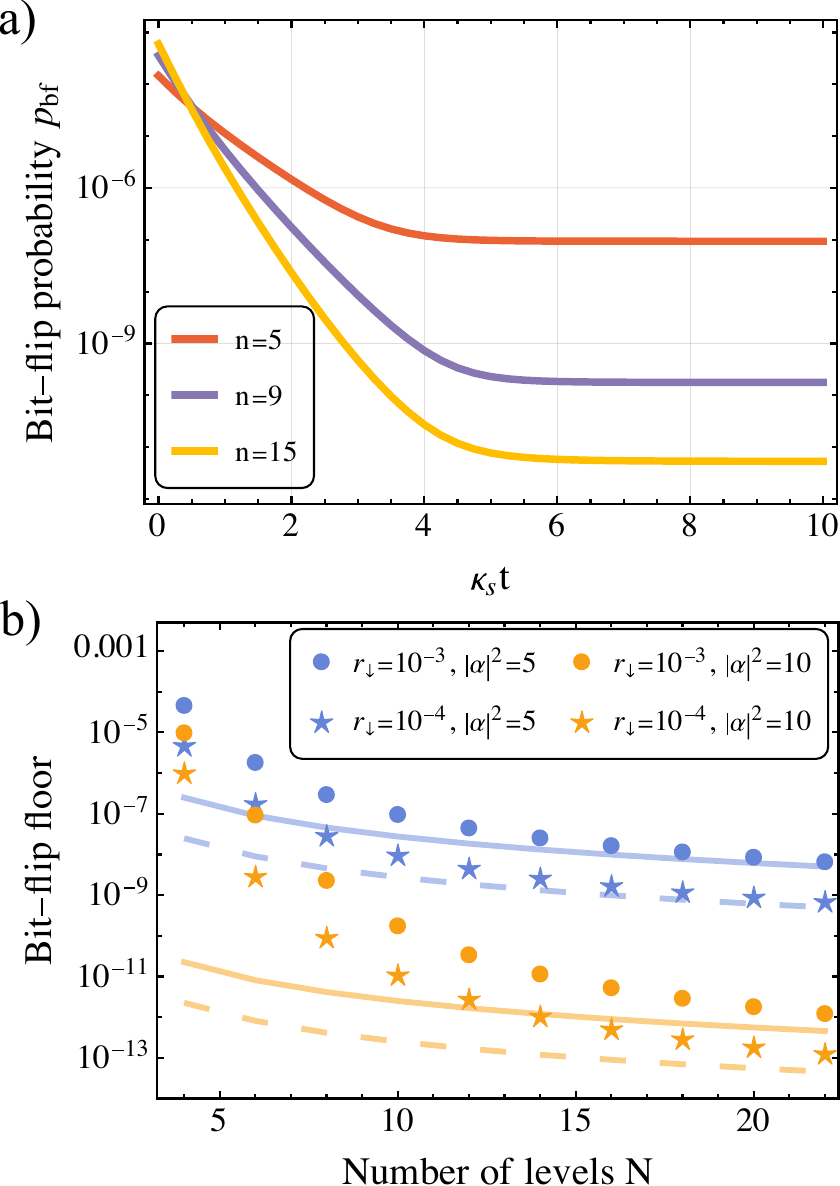}
    \caption{Bit-flip probability $p_\text{bf}$ of the CX gate on cat qubits with mean photon number $|\alpha|^2$, when the 0-n Fock qubit is used as the control qubit. $r_{\uparrow, \downarrow} = \kappa_i^{\uparrow, \downarrow}/\kappa_s$ is the ratio between the rates of the intrinsic single-photon events to the rate of engineered dissipation. For both subfigures, we set the pulse error probability as $p_e = 10^{-3}$. a) On relatively short timescales $t \sim 1/\kappa_s$, the 0-n Fock qubit stabilizes and $p_\text{bf}$ converges to a fixed value referred to as the bit-flip floor, inversely proportional to $n$. We set $|\alpha|^2 = 10$, $r_\downarrow = 10^{-3}$, and $r_\uparrow = 10^{-5} $. b) Bit-flip floor as a function of the number of levels $N=n+1$ of the Fock qubit. The floor decays exponentially with $n$ for small $n$, however the scaling is modified to $1/n^2$ for $n \gg 1$ (see Eq. (\ref{eq:bit_flip_floor_large_n})). We set $r_\downarrow=10^{-3}$ (plotted with circles) and $10^{-4}$ (plotted with stars), for $|\alpha|^2=5$ (plotted in blue) and $|\alpha|^2=10$ (plotted in yellow).  The analytical fit for large $n$ is plotted with the respective colors, with plain and dashed lines for $r_\downarrow = 10^{-3}$ and $r_\downarrow = 10^{-4}$, respectively. }
    \label{fig:bit_flip_floor}
    \sublabel{fig:bit_flip_floor_a}
\sublabel{fig:bit_flip_floor_b}
\end{figure}

First, notice that given the initial state before the stabilization in Eq. (\ref{eq:control_qubit_imperfect_pulse_final_state}), we have that
\begin{subequations}
\begin{align}
    P(0) &= \frac{1}{2}\left( p_e + (1-p_e)^n\right), \\
    P(j) &= \frac{p_e}{2}\left( 1-p_e\right)^j, \; 1 \leq j \leq n-1,\\
    P(n) &= \frac{1}{2}\left(1-p_e\right)^n.
\end{align}
\end{subequations}
Since $\kappa_i^\downarrow \gg \kappa_i^\uparrow$, the bit-flip errors will arise from intrinsic loss to lower levels in the upper half of the 0-n Fock qubit. The total population in these levels is initially $\sum_{j=(n+1)/2}^n P(j) = \frac{1}{2}(1-p_e)^{(n+1)/2}$. On the timescale $t \sim 1/\kappa_s$, the populations in the upper half will converge to the metastable state given in Eq. (\ref{eq:upper_metastable_state}) of the main text. Then, the populations in this section can be written as
\begin{align}
    P(j) &\approx \frac{(1-p_e)^{\frac{n+1}{2}}}{2}\left(\frac{\kappa_i^\downarrow}{\kappa_s +\kappa_i^\uparrow}\right)^{n-j} \nonumber \\
    &\approx \frac{(1-p_e)^{\frac{n+1}{2}}}{2} \left(\frac{\kappa_i^\downarrow}{\kappa_s }\right)^{n-j}
\end{align}
for $ \frac{n+1}{2} \leq j \leq n$. 
The level populations decay exponentially as we go to lower levels from $n$. In contrast, the bit-flip probability of a level increases. However, for large enough $n$, the decay in populations will be faster than the increase in the bit flip probability. Then, the highest contribution to the total bit-flip probability will be due to level $n-1$. The bit-flip probability can therefore be approximated as
\begin{align}
    p_\text{bf} &\approx \frac{(1-p_e)^{\frac{n+1}{2}}}{2} \left(\frac{\kappa_i^\downarrow}{\kappa_s }\right) \tilde{p}_\text{bf} \left( \pi-\frac{\pi}{n} \right)
\end{align}
where
\begin{align}
    \tilde{p}_\text{bf} \left( \pi-\frac{\pi}{n} \right)  &= \frac{1}{2} \left[ 1 - \left(1- M_\alpha \left( \pi-\frac{\pi}{n}\right)  \right)^2 \right] \nonumber \\
    &= M_\alpha\left( \frac{\pi}{n}\right) - \frac{1}{2}M_\alpha^2\left( \frac{\pi}{n}\right) 
\end{align}
as $M_\alpha(\pi - \frac{\pi}{n}) = 2-M_\alpha(\frac{\pi}{n})$. Let us approximate $M_\alpha(\frac{\pi}{n})$ for $n \gg 1$. Defining $x \coloneqq \pi/n$, $x \ll 1$, we have:
\begin{align}
    M_\alpha\left(x \right) &= \frac{2|\alpha|^2}{\sinh{(2|\alpha|^2)}} \int_0^{x} d\phi \, \sin{(\phi)} I_0\left(2|\alpha|^2\sin{(\phi)} \right) \nonumber \\
    &\approx \frac{|\alpha|^2 x^2}{\sinh{(2|\alpha|^2)}}, \; \text{for} \; x \ll 1
\end{align}
Then, for large enough $n$ such that $M_\alpha(\frac{\pi}{n}) \ll 1$, we find
\begin{align}
\label{eq:bit_flip_floor_large_n}
    p_\text{bf} &\approx \frac{\pi^2|\alpha|^2 (1-p_e)^{\frac{n+1}{2}}}{2n^2\sinh{(2|\alpha|^2)}} \left(\frac{\kappa_i^\downarrow}{\kappa_s }\right).
\end{align}
We thus observe that the bit-flip floor decays as $1/n^2$ in the limit of $n \rightarrow \infty$. This analytical expression is plotted with plain lines for $r_\downarrow = 10^{-3}$, and dashed lines for $r_\downarrow = 10^{-4}$ in Fig. \ref{fig:bit_flip_floor_b}. For both cases, blue curves represent $|\alpha|^2=5$, whereas yellow curves represent $|\alpha|^2=10$. We observe that this function is a good fit in the limit of $N \gg 1$.

\section{Analytical solution to bit-flip probability after CX gates with decay}
\label{app:analytical_CX}

For a large number of filter modes, performing the full simulation of the stabilization of the 0-n Fock qubit, as well as the noisy CX gates conditioned on it can be computationally challenging. Because of this, it is valuable to approximately quantify the bit-flip probabilities with an analytical method. We show that such a method can be obtained here. 

First, we have shown in Section \ref{sec:0_n_qubit} that the evolution of the 0-n Fock qubit during an ideal stabilization can be separated into two components: the evolution of the diagonal elements, $\ket{i}\bra{i}$, and the coherences, $\ket{i}\bra{j}$, $i\neq j$. The former can be computed with a CTMC, whereas the latter only shrinks in amplitude, where the rate can be computed
given the decay and heating rates $\kappa_i^d$, $\kappa_i^h$ between levels $\ket{j}$ and $\ket{j+1}$. These rates are obtained analytically in a perturbative fashion (cf. Eq. (\ref{eq:gaussian_approximate_rate})).

After the stabilization, 
the 0-n Fock qubit is used to perform the CX gates on the cat qubits. The interaction Hamiltonian between the Fock qubit and a cat qubit is in the form of $H = \chi \hat{n}_{a} \hat{n}_{d}$, where $\hat{n}_{x} = \hat{x}^\dagger \hat{x}$, and $\hat{a}$ ($\hat{d}$) is the annihilation operator on the Fock (cat) qubit. Furthermore, the CX gate is subject to intrinsic single-photon events. For the CX gate on $M$ cat qubits, their final state can be written as
\begin{align}
\label{eq:app_cat_final_state}
    \hat{\Psi}_f = \Phi^M(\text{Tr}_\text{F}(\Lambda^M(\hat{\rho} \otimes \hat{\Psi}_i)))
\end{align}
where $\hat{\Psi}_i \coloneqq \ket{0_L}\bra{0_L}^{\otimes M}$, $\Lambda^M(\cdot) \coloneqq \Lambda_M \circ \dots \circ\Lambda_1$, $\Phi^M(\cdot) \coloneqq \Phi_M \circ \dots \circ\Phi_1$ , $\Phi_i(\cdot) \coloneqq I^{\otimes M-i} \otimes \Phi(\cdot) \otimes I^{i-1}$, and the trace is over the Fock qubit. Furthermore, we define $\Lambda_i$ to be the channel $\Lambda$ applied to the control qubit and the $i^\text{th}$ qubit, with the identity acting on all remaining subsystems.
Here, $\Lambda$ is the CX gate, and $\Phi$ is the stabilization of the cat qubit via two-photon dissipation. Given this final state, the bit-flip probability is
\begin{align}
\label{eq:app_bf_prob}
    1-(\bra{0_L}^{\otimes M} \hat{\Psi}_f \ket{0_L}^{\otimes M} + \bra{1_L}^{\otimes M} \hat{\Psi}_f \ket{1_L}^{\otimes M}).    
\end{align}
We observe that the CX interaction satisfies
\begin{align}
    \Lambda(\ket{i}\bra{j} \otimes \hat{\rho}) \propto \ket{i}\bra{j} \otimes \hat{\sigma} \; \text{for} \; i \neq j,
\end{align}
i.e. the coherence terms of the Fock qubit are unaffected up to an amplitude rescaling due to decoherence, and the cat qubit is transformed from $\hat{\rho}$ into $\hat{\sigma}$. Due to the tracing out of the Fock qubit, these terms will not contribute to $\hat{\Psi}_f$. Furthermore, there is no cross-talk between them and the diagonal terms $\ket{i}\bra{i}$ of the Fock qubit. Then, we can ignore the coherences of the initial state of the Fock qubit while calculating the bit-flip probability.
Because of this, we work with the initial Fock qubit state 
in the form of $\hat{\rho} = \sum_j \ket{j}\bra{j}$.

\begin{figure}
    \centering
    \includegraphics[width=\linewidth]{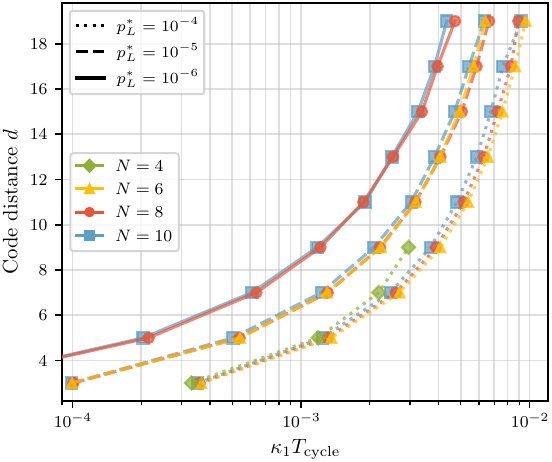}
    \caption{Average logical error per round $p_L^*$ when the 0-n Fock qubit is used as an ancilla in the repetition code, paired with cat data qubits, with a dephasing budget on the Fock qubit. The contours for $p_L^*$ are shown as a function of $\kappa_1 T_\text{cycle}$ and the code distance $d$, where
    $\kappa_1$ is the single-photon loss rate of the cat qubits, and $T_\text{cycle}$ is the cycle time for a single round of error correction. $Z$ errors are applied on the ancilla and the data qubits at the start of each error correction round. We assume single-photon events on the ancilla with $\tau_i^\downarrow = 250 \;\mu\text{s}, \,\tau_i^\uparrow = 250 $ ms, and dephasing such that the overall $Z$ error probability $p_Z^\text{ancilla}$ is doubled. Lastly, we assume a stabilization time $t_s = 500$ ns, CX gate time $t_g = 200$ ns, and cat mean photon number $|\alpha|^2=10$.}
    \label{fig:rep_code_threshold_appendix}
\end{figure}

We can then compute the effect of $\Lambda$ using the Lindbladian in Eq. (\ref{eq:lindblad_eqn}) with the CX Hamiltonian $H = \chi\hat{n}_{a} \hat{n}_{d}$. For coherent states $\ket{\alpha}, \ket{\beta}$, the Hamiltonian satisfies
\begin{align}
    [H, \ket{j}\bra{j}\otimes \ket{\alpha}\bra{\beta}] = \chi j \ket{j}\bra{j} \otimes [\hat{n}_d, \ket{\alpha}\bra{\beta}]
\end{align}
This term acts on the initial coherent state superposition to rotate them in the phase space over time, and the Fock qubit is unaffected. Conversely, the Lindbladian $\mathcal{L}(\hat{\rho})$ in Eq. (\ref{eq:lindblad_eqn}) causes single photon decay and heating events, causing transitions between adjacent Fock levels, and leaves the coherent states unchanged. Therefore, we propose the following ansatz for the evolution of the joint state of the Fock qubit $\sum_j p_j \ket{j}\bra{j}$ and the dyad $\ket{\alpha}\bra{\beta}$:
\begin{align}
    \hat{\sigma} = \sum_j \int d\theta  \, P_j(t, \theta) \ket{j}\bra{j} \otimes \ket{\alpha e^{-i\theta}}\bra{\beta e^{-i \theta}}
\end{align}
where $P_j(0, \theta) = p_j \delta(\theta) \; \forall \;j$, $\delta(x)$ is the Dirac delta function, and $\sum_j p_j = 1$. For easier notation, let us define $\alpha e^{-i\theta} = \alpha_\theta$, $\beta e^{-i\theta} = \beta_\theta$. Then, 
\begin{align}
    \frac{d\hat{\sigma}}{dt} = \sum_j  \int d\theta \, \partial_t P_j(t, \theta) \ket{j}\bra{j} \otimes \ket{\alpha_\theta}\bra{\beta_\theta}
\end{align}
Also, notice that
\begin{align}
    \frac{d \ket{\alpha_\theta}}{d \theta} &= e^{-|\alpha|^2/2} \sum_n (-i n) \frac{(\alpha \, e^{-i \theta})^n}{\sqrt{n!}}\ket{n} \nonumber \\
    &= -i \hat{n} \ket{\alpha e^{-i\theta}}
\end{align}
Then, $[\hat{n}, \ket{\alpha_\theta} \bra{\beta_\theta}] = i\frac{d}{d \theta} (\ket{\alpha_\theta} \bra{\beta_\theta})$. 
Using this identity, $-i[H, \hat{\sigma}]$ can be written as (up to the constant factor $\chi$):
\begin{align}
\label{eq:app_H}
    & \sum_j \int d\theta  \,  j P_j(t, \theta) \ket{j}\bra{j} \otimes \frac{d}{d\theta} (\ket{\alpha_\theta}\bra{\beta_\theta}) \nonumber \\
    =& - \sum_j \int d\theta  \,\partial_\theta j P_j(t, \theta) \ket{j}\bra{j} \otimes \ket{\alpha_\theta}\bra{\beta_\theta}
\end{align}
Let us also denote $\mathcal{L}(\hat{\sigma})$ with
\begin{align}
\label{eq:app_lind}
    \sum_{j,l}  \int d\theta \, M_j(l) P_j(t, \theta) \ket{j}\bra{j} \otimes \ket{\alpha_\theta}\bra{\beta_\theta}
\end{align}
The evolution of $\hat{\sigma}$ is found from
\begin{align}
    \frac{d\hat{\sigma}}{dt} = -i[H, \hat{\sigma}] + \mathcal{L}(\hat{\sigma}).
\end{align}
From Eqs. (\ref{eq:app_H}) and (\ref{eq:app_lind}), we observe this equation can be rewritten in terms of $P_j(t, \theta)$ only. Doing so, we obtain the following partial differential equation
\begin{align}
    \partial_t P_j(t, \theta) = -\chi j \partial_\theta P_j(t, \theta) + \sum_l M_j(l) P_j(t, \theta).
\end{align}
This equation can be solved by going into the Fourier domain, where $\tilde P_j(t, k) = \int d\theta e^{-i k \theta} P_j(t, \theta)$.
\begin{align}
    \partial_t \tilde P_j(t, k) &= -i\chi j k \tilde P_j(t, k) + \sum_l M_j(l) \tilde P_j(t, k) \nonumber \\
    &= \sum_l M_j'(l) \tilde P_j(t, k)
\end{align}
Solving this for each $j$ results in obtaining the evolution of $\hat{\sigma}$.
Let us refer to the solution at the end of the CX interaction as $P_j(T, \theta)$. We notice that this solution is independent of the coherent state amplitudes $\alpha, \beta$. Using this property, and given the initial state of the cat qubit before the interaction is in the form of
\begin{align}
    \ket{0_L}\bra{0_L} &=  a_{++} \ket{\alpha}\bra{\alpha} + a_{--} \ket{-\alpha}\bra{-\alpha} \nonumber \\
    &\quad + a_{+-}(\ket{\alpha}\bra{-\alpha} + \ket{-\alpha}\bra{\alpha}),
\end{align}
The final state will therefore result in
\begin{align}
    \hat{\sigma}_f = &\sum_j \int d\theta  \, P_j(T, \theta) \ket{j}\bra{j} \otimes ( a_{--}\ket{-\alpha_\theta}\bra{-\alpha_\theta} \nonumber \\ & +a_{++}\ket{\alpha_\theta}\bra{\alpha_\theta} + a_{+-} (\ket{\alpha_\theta}\bra{-\alpha_\theta} + \ket{-\alpha_\theta}\bra{\alpha_\theta} ))
\end{align}
For the two-resonator cost function considered in the main text, we need to apply the same CX interaction to a secondary cat qubit resonator initialized in $\ket{0_L}\bra{0_L}$. For an easier computation, we can commute the stabilization of the first qubit in Eq. (\ref{eq:app_cat_final_state}), as well as the projection into the logical subspace $\{\ket{0_L}, \ket{1_L}\}$, with this interaction. Thus, we obtain two possible initial Fock qubit states depending on the result of this projection. For both of these states, one can resolve the evolution for $P_j(t, \theta)$ for the second resonator, which is then stabilized and projected into the logical subspace. Finally, the bit-flip probability can be computed from Eq. (\ref{eq:app_bf_prob}) with $M=2$.

\section{Logical error rates with a dephasing budget}
\label{app:logical_error_with_dephasing}

Here, we recompute the average logical error per round, given the 0-n Fock qubit as the ancilla and cat data qubits, for the repetition code. In Sec. \ref{subsec:logical_error_rates}, we have ignored the dephasing effects as we assumed that $E_J$ can be sufficiently increased such that dephasing due to charge noise is suppressed exponentially in $E_J/E_C$. Here, we relax this assumption and work in a regime where dephasing results in a loss of coherence equal to that of the intrinsic single-photon events. In Sec. \ref{sec:0_n_qubit}, we have computed that single-photon events result in a decay of coherence terms as $\hat{\rho}(t) = e^{-\frac{1}{2}\kappa_c t} \hat{\rho}(0)$, where $\hat{\rho}(0) = \ket{m}\bra{n}$, $\kappa_c = \kappa_{m}^t+ \kappa_{n}^t$, and $\kappa_{n}^t = \kappa_{n-1}^\text{d} + \kappa_{n}^\text{h}$ is the total rate of the outgoing excitations from level $n$. Then, the coherence $\ket{0}\bra{n}$ decays with a total rate $\kappa_c = n\kappa_i^\downarrow +(n+2)\kappa_i^\uparrow \approx n\kappa_i^\downarrow$ for $\kappa_i^\downarrow \gg \kappa_i^\uparrow$. Similarly, dephasing with a Lindblad operator $\hat{L}_\phi = \sqrt{\kappa_\phi} \, \hat{a}^\dagger \hat{a}$ results in a loss of coherence with rate $\kappa_c = \kappa_\phi (m-n)^2$. Thus, we set $\kappa_\phi = \kappa_i^\downarrow/n$ for the 0-n Fock qubit, so that the decay due to dephasing is equal to that due to the single-photon loss. 

The recomputed average logical error rates (including the bit-flip contribution) can be found in Fig. \ref{fig:rep_code_threshold_appendix}. We plot the contours $p_L^* \in \{10^{-4}, 10^{-5}, 10^{-6}\}$ with dotted, dashed, and plain lines, respectively. We work with $N \in \{4,6,8,10\}$, which are represented with the different colors. We set the Fock qubit stabilization time as $t_s = 500$ ns, and the CX gate time as $t_g = 200$ ns: thus, the total $Z$ and $X$ errors on the ancilla are fixed and are independent of the cycle time $T_\text{cycle}$. $T_\text{cycle}$ determines the amount of dephasing on the data qubits, with $p_Z^\text{data} \approx |\alpha|^2 \kappa_1 T_\text{cycle}, \; |\alpha|^2 \kappa_1 T_\text{cycle} \ll 1$, as in Sec. \ref{subsec:logical_error_rates}. Here, $\kappa_1$ is the single-photon loss rate of the cat qubits. Finally, we set the mean photon number as $|\alpha|^2 = 10$.

Comparing Fig. \ref{fig:rep_code_threshold_appendix} with Fig. \ref{fig:rep_code_threshold} in Sec. \ref{subsec:logical_error_rates}, we observe that a given logical error rate for a code distance $d$ is achieved for smaller $\kappa_1 T_\text{cycle}$, i.e. for a smaller dephasing rate of the cat qubit. This is expected, as the additional dephasing budget causes an increased ancilla phase-flip probability, which deteriorates the performance of the repetition code. However, this performance change does not require an order-of-magnitude change in $\kappa_1 T_\text{cycle}$.

\bibliography{superconducting_qubits}

\end{document}